\documentclass[preprint,showpacs,preprintnumbers,amsmath,amssymb]{revtex4}

\usepackage{graphicx}% Include figure files
\usepackage{dcolumn}% Align table columns on decimal point
\usepackage{bm}% bold math
\usepackage{ifpdf}
\usepackage{epstopdf}
\usepackage{color}

\usepackage{ulem}

\newcommand{\delete}{\bgroup\markoverwith{\textcolor{red}{\rule[0.5ex]{2pt}{1pt}}}\ULon}

\begin{document}

\title{Spectroscopy of reflection-asymmetric nuclei with relativistic energy density functionals}

\author{S. Y. Xia}
\author{H. Tao}
\author{Y. Lu}
\author{Z. P. Li}\email{zpliphy@swu.edu.cn}
\affiliation{School of Physical Science and Technology, Southwest University, Chongqing 400715, China}

\author{T. Nik\v si\'c}
\author{D. Vretenar}
\affiliation{Physics Department, Faculty of Science, University of Zagreb, 10000 Zagreb, Croatia}

\date{\today}% It is always \today, today,

\begin{abstract}
Quadrupole and octupole deformation energy surfaces, low-energy excitation spectra and transition rates in fourteen isotopic chains: Xe, Ba, Ce, Nd, Sm, Gd, Rn, Ra, Th, U, Pu, Cm, Cf, and Fm, are systematically analyzed using a theoretical framework  based on a quadrupole-octupole collective Hamiltonian (QOCH), with parameters determined by constrained reflection-asymmetric and axially-symmetric relativistic mean-field calculations. The microscopic QOCH model based on 
the PC-PK1 energy density functional and $\delta$-interaction pairing is shown to accurately describe the empirical trend of low-energy quadrupole and octupole collective states, and predicted spectroscopic properties are consistent with recent microscopic calculations based on both relativistic and non-relativistic energy density functionals. Low-energy negative-parity bands, average octupole deformations, and transition rates show evidence for octupole collectivity in both mass regions, for which a microscopic mechanism is discussed in terms of evolution of single-nucleon orbitals with deformation.
\end{abstract}

%\pacs{Valid PACS appear here}

\maketitle

\section{\label{Introduction}Introduction}

Even though most deformed medium-heavy and heavy nuclei exhibit quadrupole, reflection-symmetric equilibrium shapes, there are regions in the mass table where octupole deformations (reflection-asymmetric, pear-like shapes) occur, in particular, nuclei with neutron (proton) number $N\ (Z) \approx 34, 56, 88$, and 134. Reflection-asymmetric  shapes are characterized by the occurrence of low-lying negative-parity bands, as well as pronounced electric dipole and  octupole transitions \cite{Butler96,Ahmad93,BW.15,Butler16}.  The physics of octupole correlations was extensively explored in the 1980s and 1990s (see the review of Ref. \cite{Butler96}), but there has also been a strong revival of interest in octupole shapes more recently, as shown by a number of experimental \cite{Urban12,Gaff13,Urban13,Tandel13,Spieker13,HJLi14,Scheck14,Ahmad15,Ruch15,Bucher16,Bucher17,Zimba16,Liu16} and theoretical \cite{Geng07,Moller08,Guo10,Robledo10,Robledo11,Rodr12,Robledo13,Robledo15,Bernard16,Minkov12,Jolos12,Zhao12,ZhouSG16,Nomura13,Nomura14,Nomura15,Chen15,Wang15,Bizzeti13,Agbe16,Zhang10,Zhang10b,Li13,Li16,Yao15,Zhou16,Ebata17,Yao16,Robledo12} studies. 

In a simple microscopic picture strong octupole correlations arise through a coupling of orbitals near the Fermi surface with quantum numbers 
($l$, $j$) and ($l+3$, $j+3$). This leads to reflection-asymmetric intrinsic shapes that develop either dynamically 
(octupole vibrations) or as static octupole equilibrium deformations. For instance, in the case of heavy ($Z\approx 88$ and $N\approx 134$) nuclei, the coupling of the neutron orbitals $g_{9/2}$ and $j_{15/2}$, and that of the proton single-particle states $f_{7/2}$ and $i_{13/2}$, can lead to octupole mean-field deformations. In particular, evidence for pronounced octupole deformation in  $^{224}$Ra \cite{Gaff13}, $^{144}$Ba \cite{Bucher16}, and $^{146}$Ba \cite{Bucher17} has recently been reported in Coulomb excitation experiments with radioactive ion beams. The renewed interest in studies of reflection-asymmetric nuclear shapes using accelerated radioactive beams point to the importance of a timely systematic theoretical analysis of quadrupole-octupole collective states of nuclei in different mass regions.

Coexistence of different shapes, and shape transitions as a function of nucleon number, present universal phenomena that occur in light, medium-heavy and heavy nuclei. A unified description of the evolution of quadrupole and octupole states necessitates a universal theory framework that can be applied to different mass regions.
Nuclear energy density functionals (EDFs), in particular, enable a complete and accurate description of ground-state properties and collective excitations over the entire chart of nuclides \cite{Bender03,VALR05,Meng06,Stone07,Nik11,Meng16}. Both non-relativistic and relativistic EDFs have successfully been applied to the description of the evolution of single-nucleon shell structures and related nuclear shapes and shape transitions.  In the literature one finds a number of detailed self-consistent mean-field studies of nuclei with static or dynamic octupole deformations, e.g., based on the Skyrme \cite{Bonche86,Bonche88} and Gogny \cite{Egido91,Robledo10,Robledo11,Rodr12,Robledo13,Robledo15,Bernard16,Robledo12} effective interactions, and relativistic mean-field (RMF) models \cite{Geng07,Guo10,Zhao12,Nomura13,Nomura14,Zhang10,Zhang10b,Li13,Li16,Yao15,Zhou16}. To compute excitation spectra and transition rates, however, the EDF framework has to be extended to take into account the restoration of symmetries broken in the mean-field approximation, and fluctuations in the collective coordinates. A straightforward approach is the generator coordinate method (GCM) combined with projection techniques, and recently it has been implemented for reflection-asymmetric shapes, based on both nonrelativistic \cite{Bernard16} and relativistic \cite{Yao15,Zhou16} EDFs. Using this method, however, it is rather difficult to perform a systematic study of low-lying quadrupole and octupole states in different mass regions, because implementations of GCM are very time-consuming for heavy systems.  Possible alternative approaches are the EDF-based interacting boson model \cite{Nomura13,Nomura14,Nomura15}, or the quadrupole-octupole collective Hamiltonian \cite{Li13,Li16}. In particular, the EDF-based collective Hamiltonian can be derived from the GCM in the Gaussian overlap approximation \cite{Ring80}, and the validity of this approximate method was recently demonstrated in a comparison with a full GCM calculation for the shape coexisting nucleus $^{76}$Kr \cite{Yao14}. 

In this study we employ the recently developed  EDF-based quadrupole-octupole collective Hamiltonian (QOCH) \cite{Li13,Li16} to perform a systematic calculation of even-even medium-heavy ($54\leq Z\leq 64$ and $84\leq N\leq100$), and heavy nuclei ($86\leq Z\leq 100$ and $130\leq N\leq152$). Low-energy spectra and transition rates for both positive- and negative-parity states of 150 nuclei are calculated using the QOCH with parameters determined by self-consistent reflection-asymmetric relativistic mean-field calculations based on the PC-PK1 energy density functional \cite{Zhao10}. The relativistic functional PC-PK1 was adjusted to the experimental masses of a set of 60 spherical nuclei along isotopic or isotonic chains, and to the charge radii of 17 spherical nuclei.  PC-PK1 has been successfully employed in studies of nuclear masses \cite{Zhang14,Lu15}, and spectroscopy of low-lying quadrupole states \cite{Quan17}.

The article is organized as follows. Sec. \ref{ScII} describes the theoretical framework, and an illustrative calculation of $^{224}$Ra is presented in Sec. \ref{IIIA}. The systematics of collective deformation energy surfaces, excitation energies of low-lying positive- and negative-parity states, electric dipole, quadrupole, and octupole transition rates, calculated with the EDF-based QOCH, are discussed in Sec. \ref{IIIB}. Section \ref{IV} contains a summary and concluding remarks.

%%%%%%%%%%%%%%%%%%%%%%%%%%%%%%%%%%%%%%%%%%%%%%%%%%%%%%
%
\section{\label{ScII}Theoretical Framework}
%
%%%%%%%%%%%%%%%%%%%%%%%%%%%%%%%%%%%%%%%%%%%%%%%%%%%%%%
 %
 \subsection{The quadrupole-octupole collective Hamiltonian}

Nuclear excitations characterized by quadrupole and octupole vibrational and rotational degrees of freedom can be simultaneously described by considering quadrupole and octupole collective coordinates that specify the surface of a nucleus $R=R_0\left[1+\sum_\mu{\alpha_{2\mu}Y^*_{2\mu}+ \sum_\mu{\alpha_{3\mu}Y_{3\mu}^*} } \right]$. In addition, when axial symmetry is imposed, the collective coordinates can be parameterized in terms of two deformation parameters $\beta_2$ and $\beta_3$, and three Euler angles  $ \Omega\equiv(\phi,\theta,\psi)$:
 \begin{equation} 
  \alpha_{\lambda \mu} = \beta_\lambda D_{0\mu}^\lambda(\Omega), \quad \lambda=2,3 .
\end{equation}
The classical collective Hamiltonian is expressed as the sum of the vibrational kinetic energy, rotational kinetic energy, and the collective potential $\mathcal{V}_{\rm coll}$.  The vibrational and rotational kinetic energies read:
\begin{eqnarray}
{\cal T}_{\rm vib}&=&\frac{1}{2}B_{22}\dot{\beta}_2^2+B_{23}\dot{\beta}_2\dot{\beta}_3
+\frac{1}{2}B_{33}\dot{\beta}_3^2, \\
{\cal T}_{\rm rot} &=& \frac{1}{2}\sum\limits_{k=1}^3{\cal I}_k\omega_k^2,
\end{eqnarray} 
respectively, where the mass parameters $B_{22}$, $B_{23}$ and $B_{33}$, and the moments of inertia $\mathcal{I}_k$, are functions of the quadrupole and octupole deformations $\beta_2$ and $\beta_3$.

After quantization the collective Hamiltonian takes the form
\begin{eqnarray}
{\hat H}_{\rm coll} &=& -\frac{\hbar^2}{2\sqrt{w{\cal I}}}
              \left[\frac{\partial}{\partial\beta_2}\sqrt{\frac{{\cal I}}{w}}B_{33}\frac{\partial}{\partial\beta_2}
                      -\frac{\partial}{\partial\beta_2}\sqrt{\frac{{\cal I}}{w}}B_{23}\frac{\partial}{\partial\beta_3}\right.\nonumber\\
     & &
             \left. -\frac{\partial}{\partial\beta_3}\sqrt{\frac{{\cal I}}{w}}B_{23}\frac{\partial}{\partial\beta_2}
                      +\frac{\partial}{\partial\beta_3}\sqrt{\frac{{\cal I}}{w}}B_{22}\frac{\partial}{\partial\beta_3}\right]\nonumber\\
     & &
           +\frac{\hat{J}^2}{2{\cal I}}+{V}_{\rm coll}(\beta_2, \beta_3),
   \label{eq:CH2}
\end{eqnarray}
where $w=B_{22}B_{33}-B_{23}^2$ and the corresponding volume element in the collective space reads
\begin{equation}
\int d\tau_{\rm coll} = \int\sqrt{w{\cal I}}d\beta_2 d\beta_3 d\Omega.
\end{equation}

To solve the eigenvalue problem of the collective Hamiltonian Eq.~$(\ref{eq:CH2})$, an expansion of eigenfunctions in terms of a complete set of basis functions is employed. For each value of the angular momentum $I$ the basis is defined by the following relation:
\begin{equation}
\label{eq:basis}
|n_2n_3IMK\rangle=(w{\cal I})^{-1/4}\phi_{n_2}(\beta_2)\phi_{n_3}(\beta_3)|IMK\rangle,
\end{equation}
where $\phi_{n_2}$ ($\phi_{n_3}$) denotes the one-dimensional harmonic oscillator wave function of $\beta_2$ ($\beta_3$). For positive (negative) parity states, $n_3$ and $I$ are even (odd) numbers. Since we consider only axially deformed shapes, the intrinsic projection of the total angular momentum $K=0$. Finally, the collective wave function can be written as
\begin{equation}\label{collwavefunc}
\Psi^{IM\pi}_\alpha(\beta_2, \beta_3, \Omega)=\psi^{I\pi}_\alpha(\beta_2, \beta_3)|IM0\rangle,
\end{equation}
and the corresponding probability density distribution reads
\begin{equation}
\rho^{I\pi}_\alpha(\beta_2, \beta_3)=\sqrt{w{\cal I}}|\psi^{I\pi}_\alpha(\beta_2, \beta_3)|^2,
\label{eq:rho}
\end{equation}
with the normalization 
\begin{equation}
\int\rho^{I\pi}_\alpha(\beta_2, \beta_3)d\beta_2 d\beta_3=1.
\end{equation}

The reduced $E\lambda$ values are calculated from the relation 
\begin{eqnarray} \nonumber
&{}&B(E\lambda, I_i\to I_f)=\\
&{}&\langle I_i0\lambda0|I_f0\rangle^2\left|\int d\beta_2 d\beta_3 \sqrt{w{\cal I}} \psi_i
         \mathcal{M}_{E\lambda}(\beta_2,\beta_3) \psi^*_{f}\right|^2\; ,
\end{eqnarray}
where $\mathcal{M}_{E\lambda}(\beta_2,\beta_3)$ denotes the electric moment of order $\lambda$. In microscopic models it is calculated as   
$\langle\Phi(\beta_2, \beta_3)|\hat{\mathcal{M}}(E\lambda)|\Phi(\beta_2, \beta_3)\rangle$,  where $\Phi(\beta_2, \beta_3)$ is the nuclear wave function. For the electric dipole, quadrupole, and octupole transitions, the corresponding operators $\hat{\mathcal{M}}(E\lambda)$ read
\begin{eqnarray}
  D_1       &=& \sqrt{\frac{3}{4\pi}}e\left(\frac{N}{A}z_p-\frac{Z}{A}z_n\right) \label{eq:D1}\\
  Q^p_2 &=& \sqrt{\frac{5}{16\pi}}e\left(2z_p^2-x_p^2-y_p^2\right)   \label{eq:Q2p} \\
  Q^p_3 &=& \sqrt{\frac{7}{16\pi}}e\left[2z_p^3-3z_p(x_p^2+y_p^2)\right]  \label{eq:Q3p} \;,
\end{eqnarray}
respectively. 

%%%%%%%%%%%%%%%%%%%%%%%%%%%%%%%%
\subsection{Parameters of the collective Hamiltonian}

The entire dynamics of the collective Hamiltonian Eq.~(\ref{eq:CH2}) is governed by five functions of the intrinsic deformations $\beta_2$ and $\beta_3$: the collective potential, the three mass parameters $B_{22}$, $B_{23}$, $B_{33}$, and the moment of inertia $\mathcal{I}$. These functions are determined by constrained self-consistent mean-field calculations for a specific choice of the nuclear energy density functional and pairing interaction. In the present study the energy density functional PC-PK1 \cite{Zhao10} determines the effective interaction in the particle-hole channel, and in the particle-particle channel we use a $\delta$-force: $V(\mathbf{r}, \mathbf{r}^\prime)=V_{p,n}\delta(\mathbf{r}-\mathbf{r}^\prime)$, where $V_{p,n}$ are the pairing strengths for protons and neutrons, respectively \cite{Bender00}.

The entire map of the energy surface as function of the quadrupole and octupole deformations is obtained by imposing constraints on the quadrupole and octupole mass moments, respectively.  The method of quadratic constraints uses an unrestricted variation of the function
\begin{equation}
\langle H\rangle
   +\sum_{\lambda=2,3}{C_{\lambda}\left(\langle \hat{Q}_{\lambda}  \rangle - q_{\lambda}  \right)^2} \; ,
\label{constr}
\end{equation}
where $\langle H\rangle$ is the total energy, and  $\langle \hat{Q}_{\lambda}\rangle$  denotes the expectation value of the mass quadrupole and octupole operators:
\begin{equation}
\hat{Q}_{2}=2z^2-x^2-y^2 \quad \textnormal{and}\quad \hat{Q}_{3}=2z^3-3z(x^2+y^2) \;.
\end{equation}
$q_{\lambda}$ is the constrained value of the multipole moment, and $C_{\lambda}$ the corresponding stiffness constant~\cite{Ring80}. The corresponding deformation parameters $\beta_2$ and $\beta_3$ are determined from the following relations:
\begin{eqnarray}
\beta_2 &=& \frac{\sqrt{5\pi}}{3AR_0^2}\langle \hat Q_2\rangle , \\
\beta_3 &=& \frac{\sqrt{7\pi}}{3AR_0^3}\langle \hat Q_3\rangle ,
\end{eqnarray}
with $R_0=r_0A^{1/3}$ and $r_0=1.2$ fm.

The single-nucleon wave functions, energies, and occupation factors, generated from constrained self-consistent solutions of the relativistic mean-field  plus BCS-pairing equations (RMF+BCS), provide the microscopic input for the parameters of the collective Hamiltonian.

The moments of inertia are calculated according to the Inglis-Belyaev formula: \cite{Inglis56,Belyaev61}
\begin{equation}
\label{eq:MOI}
\mathcal{I} = \sum_{i,j}{\frac{\left(u_iv_j-v_iu_j \right)^2}{E_i+E_j}
  | \langle i |\hat{J} | j  \rangle |^2},
\end{equation}
where $\hat J$ is the angular momentum along the axis perpendicular to the symmetric axis,  and the summation runs over the proton and neutron quasiparticle states. The quasiparticle energies $E_i$, occupation probabilities $v_i$, and single-nucleon wave functions $\psi_i$ are determined by solutions of the constrained RMF+BCS equations. The mass parameters associated with $q_2=\langle\hat{Q}_{2}\rangle$ and $q_3=\langle\hat{Q}_{3}\rangle$ are calculated in the perturbative  cranking approximation \cite{Girod79}
\begin{equation}
\label{eq:}
B_{\lambda\lambda^\prime}(q_2,q_3)=\frac{\hbar^2}{2}
 \left[\mathcal{M}_{(1)}^{-1} \mathcal{M}_{(3)} \mathcal{M}_{(1)}^{-1}\right]_{\lambda\lambda^\prime}\;,
\end{equation}
with
\begin{equation}
\label{masspar-M}
\mathcal{M}_{(n),\lambda\lambda^\prime}(q_2,q_3)=\sum_{i,j}
 {\frac{\left\langle i\right|\hat{Q}_{\lambda}\left| j\right\rangle
 \left\langle j\right|\hat{Q}_{\lambda^\prime}\left| i\right\rangle}
 {(E_i+E_j)^n}\left(u_i v_j+ v_i u_j \right)^2}\;.
\end{equation}

The deformation energy surface (DES) includes the energy of zero-point motion that has to be subtracted. The vibrational and rotational zero-point energy (ZPE) corrections are calculated in the cranking approximation \cite{Girod79,Ni09,Li09a}:
\begin{equation}
\label{ZPE-vib}
\Delta V_{\rm vib}(\beta_2, \beta_3) = \frac{1}{4}
\textnormal{Tr}\left[\mathcal{M}_{(3)}^{-1}\mathcal{M}_{(2)}  \right]\;,
\end{equation}
and
\begin{equation}
\label{ZPE-rot}
\Delta V_{\rm rot}(\beta_2, \beta_3)=\frac{\langle\hat J^2\rangle}{2{\cal I}}\;,
\end{equation}
respectively. The potential $V_{\rm coll}$ in the collective Hamiltonian (\ref{eq:CH2}) is obtained by subtracting the ZPE corrections from the total mean-field energy:
\begin{equation}
\label{eq:Vcoll}
V_{\rm coll}(\beta_2, \beta_3)=E_{\rm tot}(\beta_2, \beta_3)-\Delta V_{\rm vib}(\beta_2, \beta_3)-\Delta V_{\rm rot}(\beta_2, \beta_3).
\end{equation}

%%%%%%%%%%%%%%%%%%%%%%%%%%%%%%%%%%%
\section{Results and discussion}

\subsection{\label{IIIA}Illustrative study of $^{224}$Ra}

\begin{figure}[ht]
\begin{center}
\includegraphics[height=0.6\textwidth]{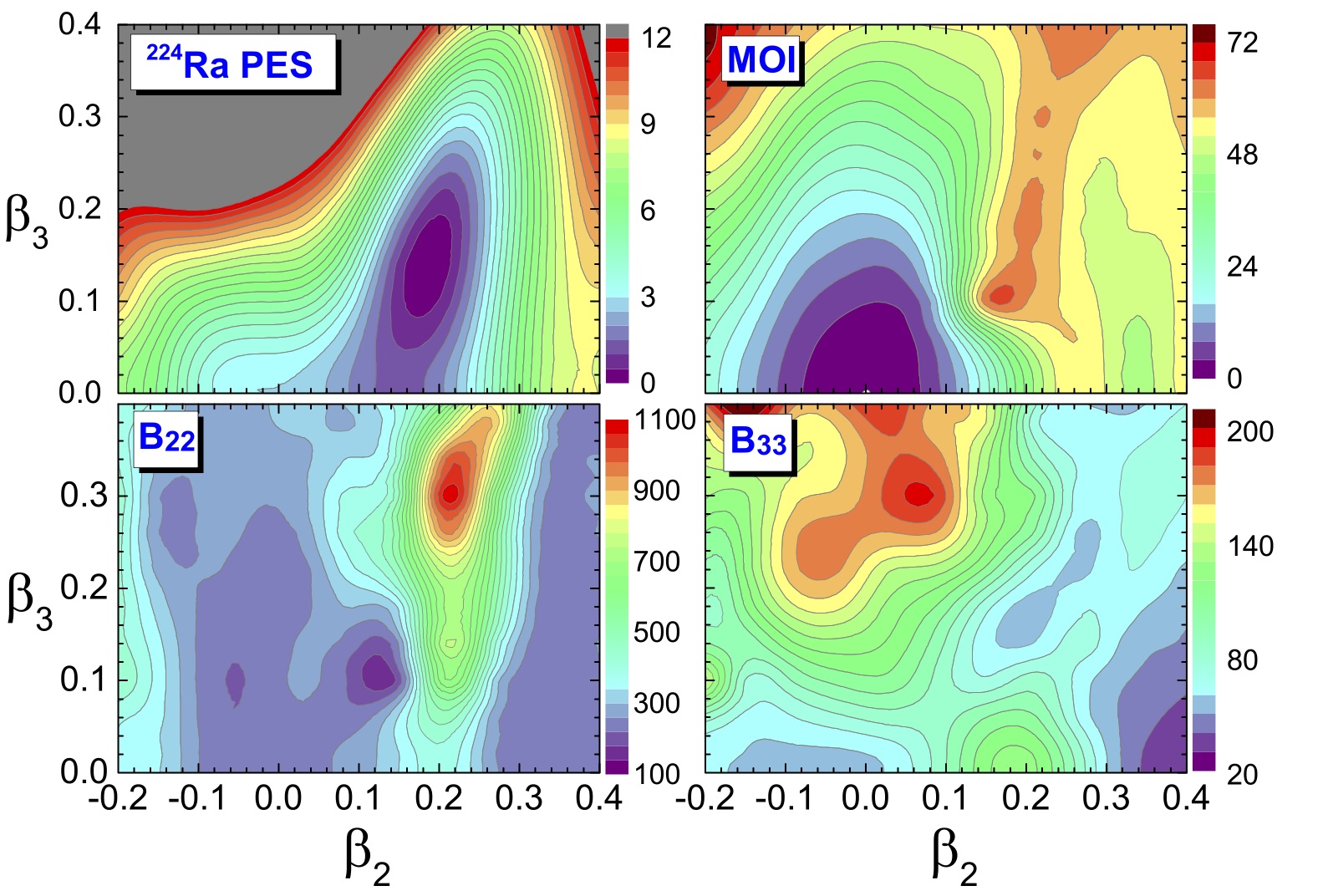}
\caption{(Color online) The deformation energy surface, moment of inertia, and the collective masses $B_{22}$ and $B_{33}$ of $^{224}$Ra in the $\beta_2-\beta_3$ plane, calculated with the RMF+BCS model.}
\label{ra224pes}
\end{center}
\end{figure}

\begin{figure}[ht]
\begin{center}
\includegraphics[height=0.7\textwidth]{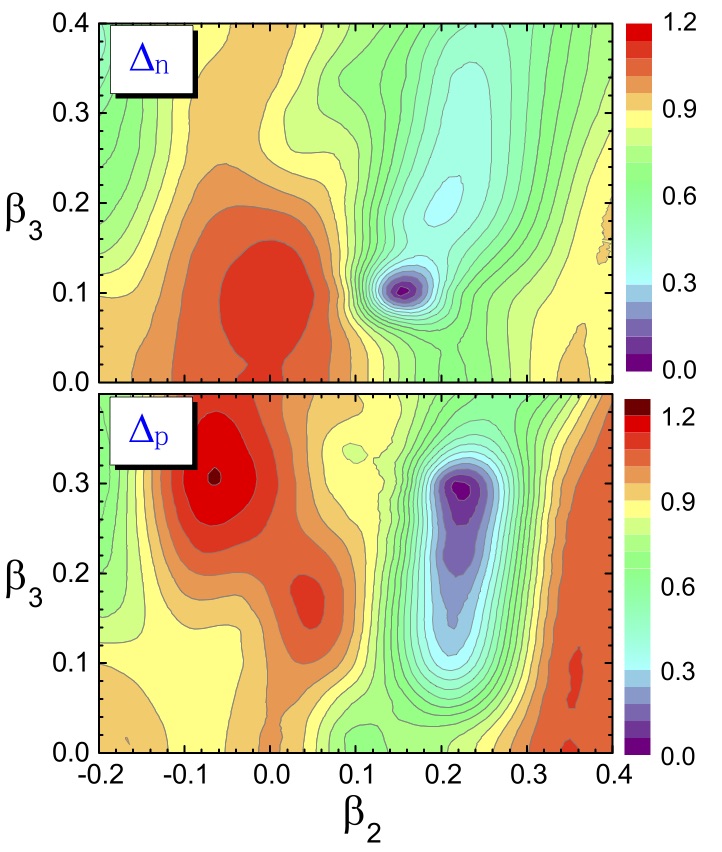}
\caption{(Color online) Neutron and proton pairing gaps (in the unit of MeV) of $^{224}$Ra as functions of deformation in the $\beta_2-\beta_3$ plane.}
\label{ra224del}
\end{center}
\end{figure}

As an illustrative example, the EDF-based quadrupole-octupole collective Hamiltonian (QOCH) is used to calculate the low-energy excitation spectrum and transitions of $^{224}$Ra. To determine the microscopic input for the QOCH, we perform a constrained reflection-asymmetric RMF+BCS calculation, with the effective interaction in the particle-hole channel defined by the relativistic point-coupling density functional PC-PK1, and a density independent $\delta$-force is the effective interaction in the particle-particle channel. The strength parameter of the $\delta$-force: $V_n=319.0$ MeV fm$^3$ ($V_p=358.3$ MeV fm$^3$) for neutrons (protons), is determined to reproduce the corresponding pairing gap of the spherical configuration of $^{224}$Ra, calculated using the relativistic Hartree-Bogoliubov (RHB) model with the finite-range separable pairing force \cite{Nik14}. The RHB model with the finite-range separable pairing force was successfully used in the description of octupole deformations \cite{Agbe16} and low-energy excitation spectra \cite{Nomura14}. For the systematic calculations reported in Sec. \ref{IIIB}, the same pairing strengths of the $\delta$-force are used for heavy nuclei with $86\leq Z\leq 100$, whereas for the medium-heavy nuclei with $54\leq Z\leq 64$ the strength parameters $V_n (V_p)=353.0 (367.0)$ MeV fm$^3$ have been determined by adjusting to the pairing gaps of the spherical configuration of $^{144}$Ba.

\begin{figure*}[ht]
\includegraphics[height=0.75\textwidth]{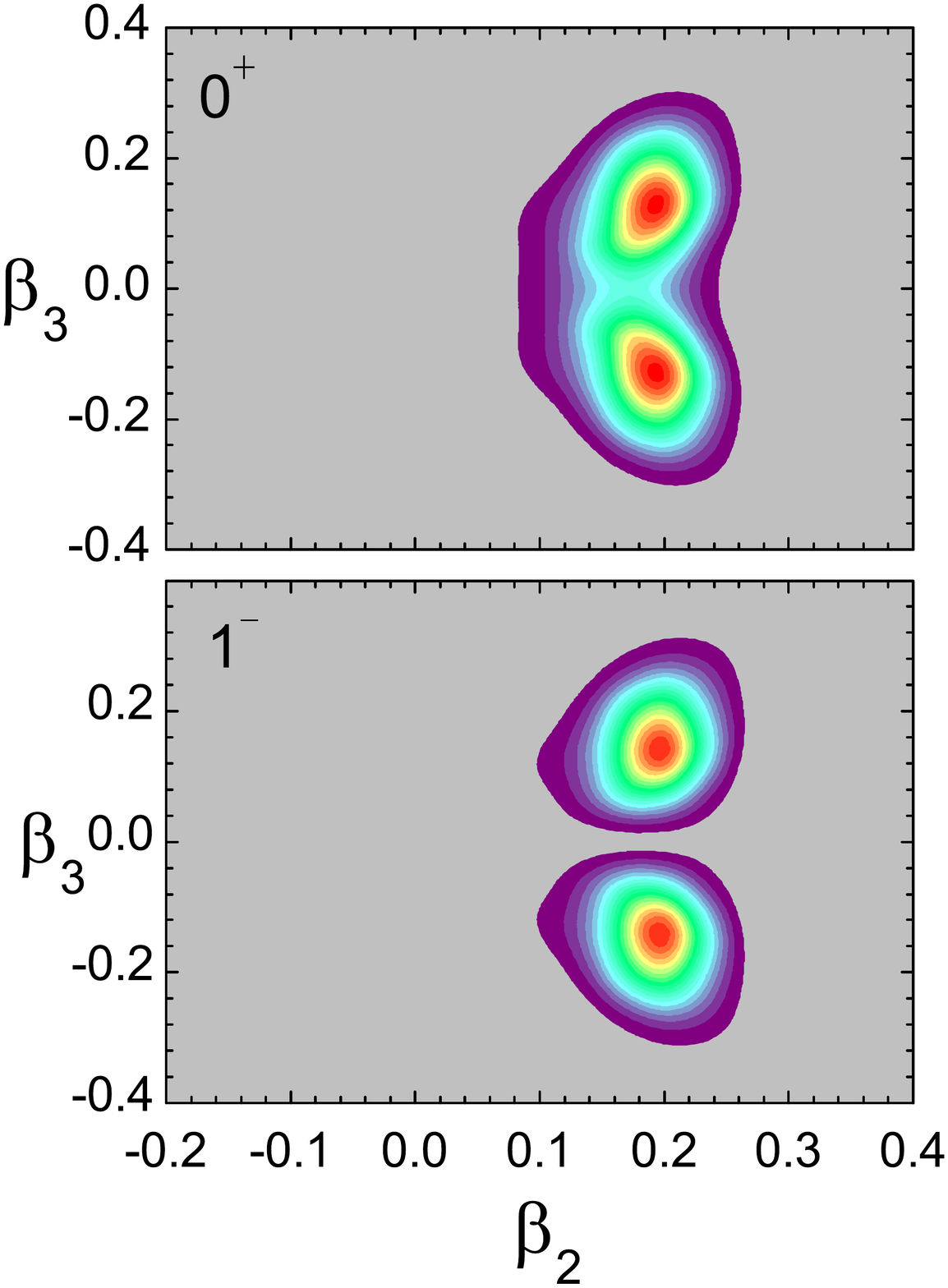}
\caption{(Color online) Probability density distributions $\rho^{I\pi}_\alpha(\beta_2, \beta_3)$ for the ground state $0^+$, and the first negative-parity state $1^-$
of $^{224}$Ra, in the $\beta_2-\beta_3$ deformation plane.}
\label{ra224wave}
\end{figure*}

Figure \ref{ra224pes} displays the deformation energy surface (DES), moment of inertia, and the collective masses $B_{22}$ and $B_{33}$ of $^{224}$Ra in the $\beta_2-\beta_3$ plane, obtained by imposing constraints on the expectation values of the quadrupole moment $\langle \hat{Q}_{2}\rangle$  and octupole moment $\langle \hat{Q}_{3}\rangle$. The DES exhibits a global minimum at $(\beta_2, \beta_3)=(0.18, 0.14)$, and is rather soft along the octupole direction. Similar patterns are also predicted by the RHB calculation with the DD-PC1 functional \cite{Nomura14},  and the  Hartree-Fock-Bogoliubov (HFB) calculation with the Gogny D1S and D1M forces \cite{Robledo13}. Generally, the moment of inertia increases with deformation. The mass parameters, on the other hand, display a more complex dependence on $\beta_2$ and $\beta_3$, caused by the fluctuations of pairing correlations.  In Fig.~\ref{ra224del} we plot the contour maps of the neutron and proton pairing gaps in the $\beta_2-\beta_3$ plane. The fluctuations of pairing gaps reflect the underlying shell structure, and pairing is strongly reduced wherever the level density around the Fermi level is small.  As a result, mass parameters are locally enhanced in regions of weak pairing. For instance, the anomalous value of $B_{22}$ at $(\beta_2, \beta_3)\sim(0.22, 0.30)$ can be related to the collapse of proton pairing at this deformation. To avoid such anomalous behavior of collective mass, in other words to avoid pairing collapse, one can perform particle number projection \cite{Egido82,Angu01}. However, this procedure is outside the scope of the present study. It is, in fact, found that the anomalous value of $B_{22}$ has negligible effect on the low-lying spectrum of $^{224}$Ra because the corresponding deformed configuration contributes very little to the total collective wave function (cf. Fig. \ref{ra224wave}).

\begin{figure*}[ht]
\includegraphics[height=0.54\textwidth]{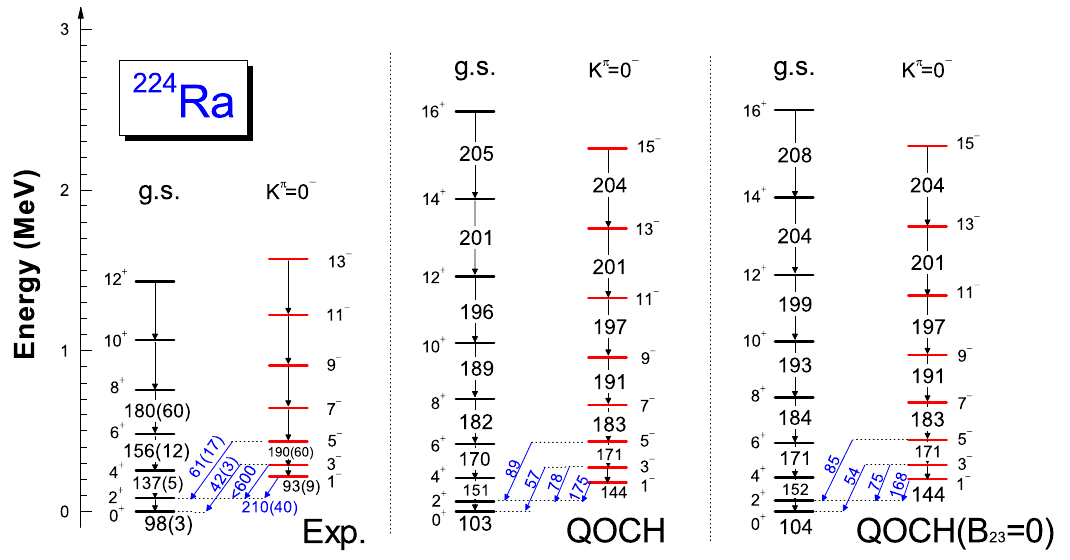}
\caption{(Color online) The excitation spectrum, intraband B(E2) (in W.u.), and interband B(E3) (in W.u.) values calculated  with the QOCH based on PC-PK1 relativistic density functional (middle), compared to experimental results (left) \cite{Gaff13}. The results obtained using the QOCH with $B_{23}=0$ in Eq. (\ref{eq:CH2}) are also shown in the right column.}
\label{ra224spec}
\end{figure*}

\begin{figure}[ht]
\begin{center}
\includegraphics[height=0.75\textwidth]{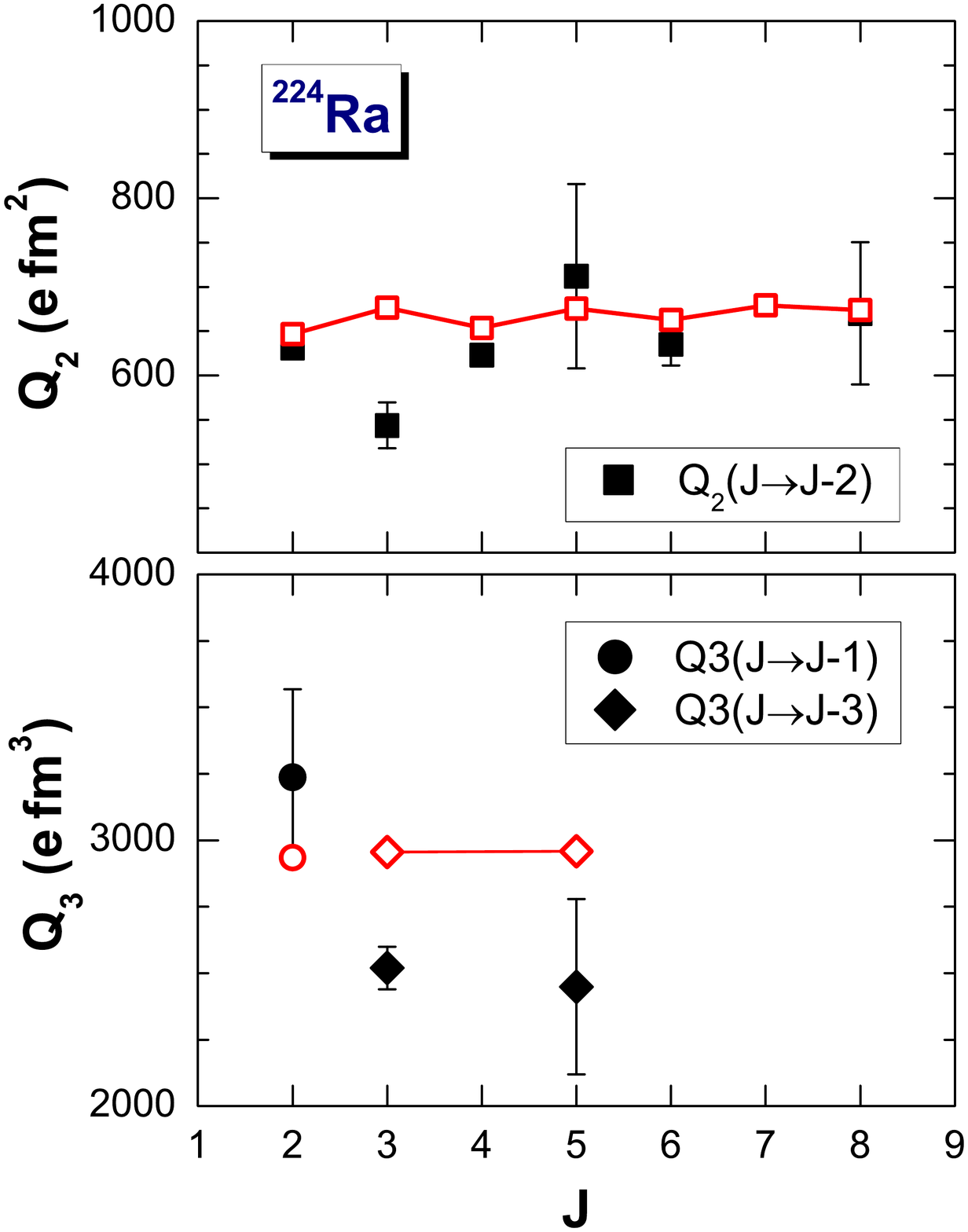}
\caption{(Color online) Comparison between the calculated (open symbols) and experimental (solid symbols) 
quadrupole and octupole intrinsic moments of $^{224}$Ra, as functions of the angular momentum.}
\label{ra224QI}
\end{center}
\end{figure}

In Fig.~\ref{ra224spec}  we compare the low-energy excitation spectrum of positive- and negative-parity states, the corresponding $B(E2)$ values for intraband transitions, and the interband $B(E3)$ values calculated with the QOCH based on PC-PK1 relativistic density functional, with recent data for the octupole deformed nucleus $^{224}$Ra obtained in the Coulomb excitation experiment of Ref.~\cite{Gaff13}.  To illustrate the effect of dynamical quadrupole-octupole coupling, we also plot the theoretical results obtained with the QOCH by setting the collective mass $B_{23}=0$ in Eq. (\ref{eq:CH2}). The difference between the two calculations is indeed very small, indicating that the dynamical coupling in this nucleus is rather weak. The level scheme of $^{224}$Ra shows that the lowest negative-parity band is located close in energy to the ground-state positive-parity band. One notices that the lowest positive- and negative-parity bands form a single, alternating-parity band,  starting with angular momentum $J=5$. Overall, a good agreement between theory and experiment is obtained for the excitation spectrum of $^{224}$Ra. The calculated $E2$ and $E3$ transition rates are also in reasonable agreement with the experimental  values. Furthermore, in Fig.~\ref{ra224QI} we plot the $E2$ and $E3$ intrinsic moments determined from the corresponding $B(E2)$ and $B(E3)$ values, respectively, using the relation $Q_\lambda(J\to J^\prime)=\sqrt{\frac{16\pi}{2\lambda+1}\frac{B(E\lambda; J\to J^\prime)}{(J\lambda 00|J^\prime 0)}}$. One notes a weak staggering in the calculated $Q_2(J\to J-2)$ values, and their average value $\sim 650$ $e~{\rm fm}^2$ is consistent with the measured value $632\pm10$ $e~\rm{fm}^2$ \cite{Gaff13}. The three theoretical  $Q_3$ moments are almost constant, $\sim2950$ $e~{\rm fm}^3$, while they underestimate the measured $Q_3(J\to J-1)$ and overestimate the $Q_3(J\to J-3)$.

Figure~\ref{ra224wave} displays the probability density distributions $\rho^{I\pi}_\alpha(\beta_2, \beta_3)$ defined in Eq. (\ref{eq:rho}), for the ground state $0^+$ and the first negative-parity state $1^-$. The distributions are, of course, symmetric with respect to $\beta_3$. A peak at $(\beta_2, \beta_3)\sim(0.18, 0.14)$ for the ground state is consistent with the global minimum of the DES, as shown in Fig.~\ref{ra224pes}. The probability density distribution of  the first negative-parity state is similar to that of the ground state, except for the symmetry requirement that the collective wave function is zero along the $\beta_3 = 0$ line. The peaks of the $1^{-}$ state are calculated at slightly larger $|\beta_3|$ compared to those of the ground state.

\clearpage
%------------------------------------------------------------------------------------
\subsection{\label{IIIB} Systematics of quadrupole and octupole states}

\begin{figure}[ht]
\includegraphics[height=0.53\textwidth]{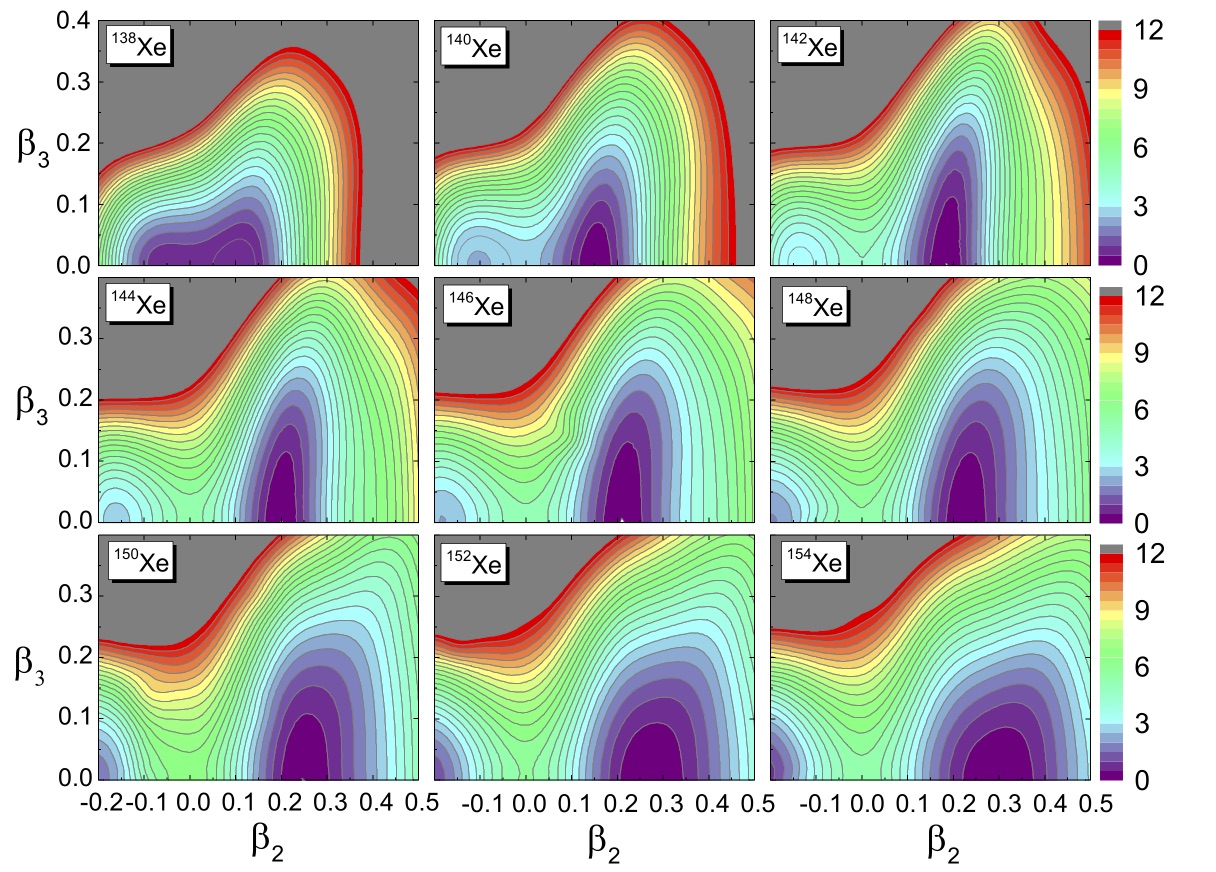}
\includegraphics[height=0.53\textwidth]{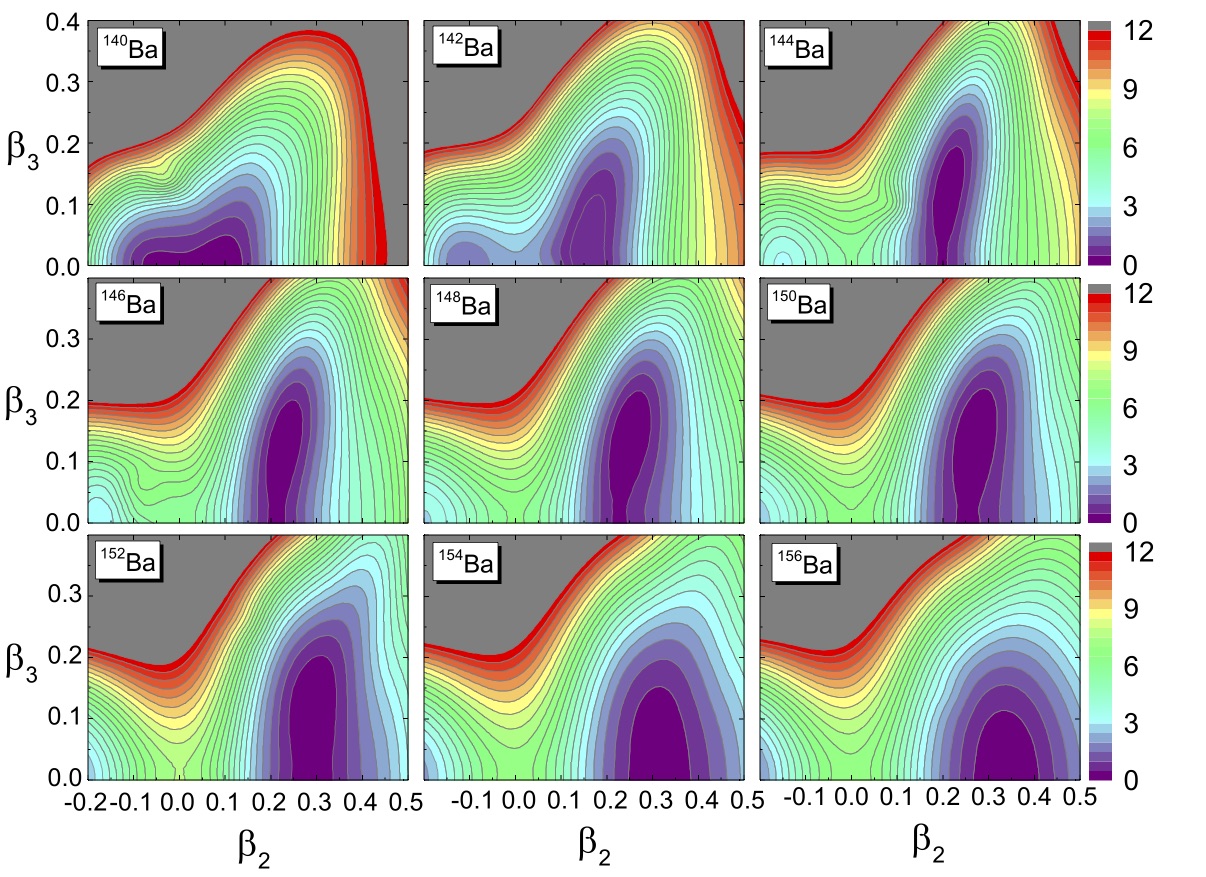}
\caption{(Color online) Deformation energy surfaces of $^{138-154}$Xe and $^{140-156}$Ba in the $\beta_2-\beta_3$ deformation plane, calculated with the RMF+BCS model using the PC-PK1 functional and $\delta$-force pairing.}
\label{PESXeBa}
\end{figure}

\begin{figure}[ht]
\includegraphics[height=0.53\textwidth]{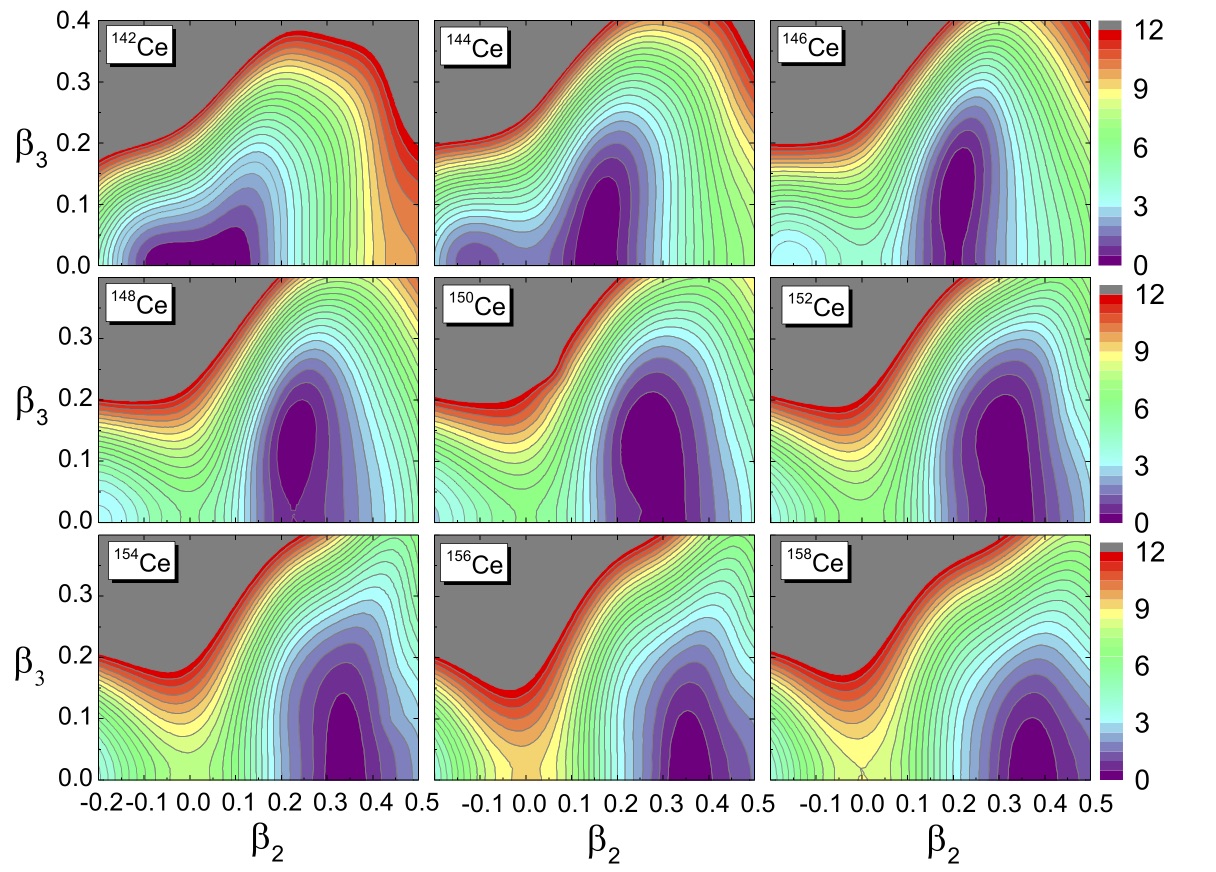}
\includegraphics[height=0.53\textwidth]{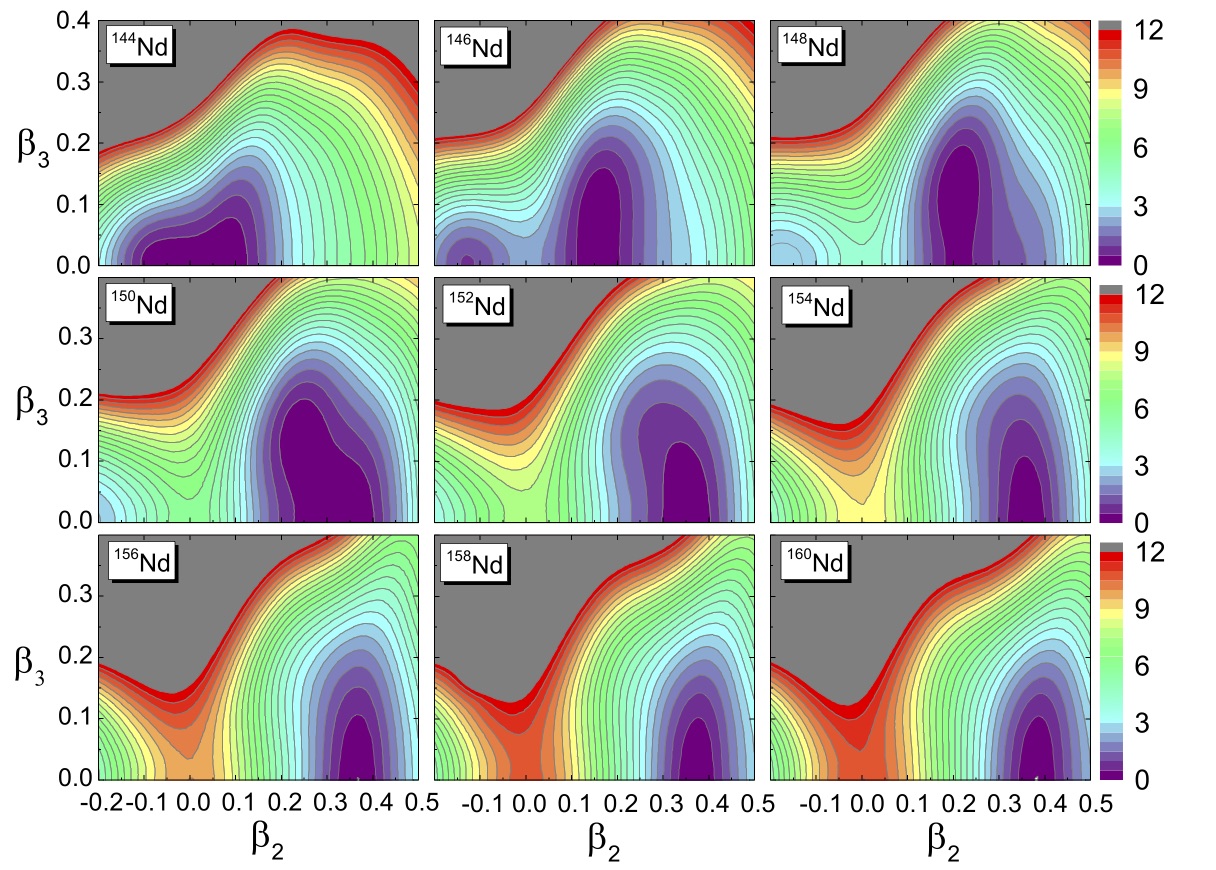}
\caption{(Color online) Same as in the caption to Fig. \ref{PESXeBa} but for the isotopes of Ce and Nd.}
\label{PESCeNd}
\end{figure}

\begin{figure}[ht]
\includegraphics[height=0.53\textwidth]{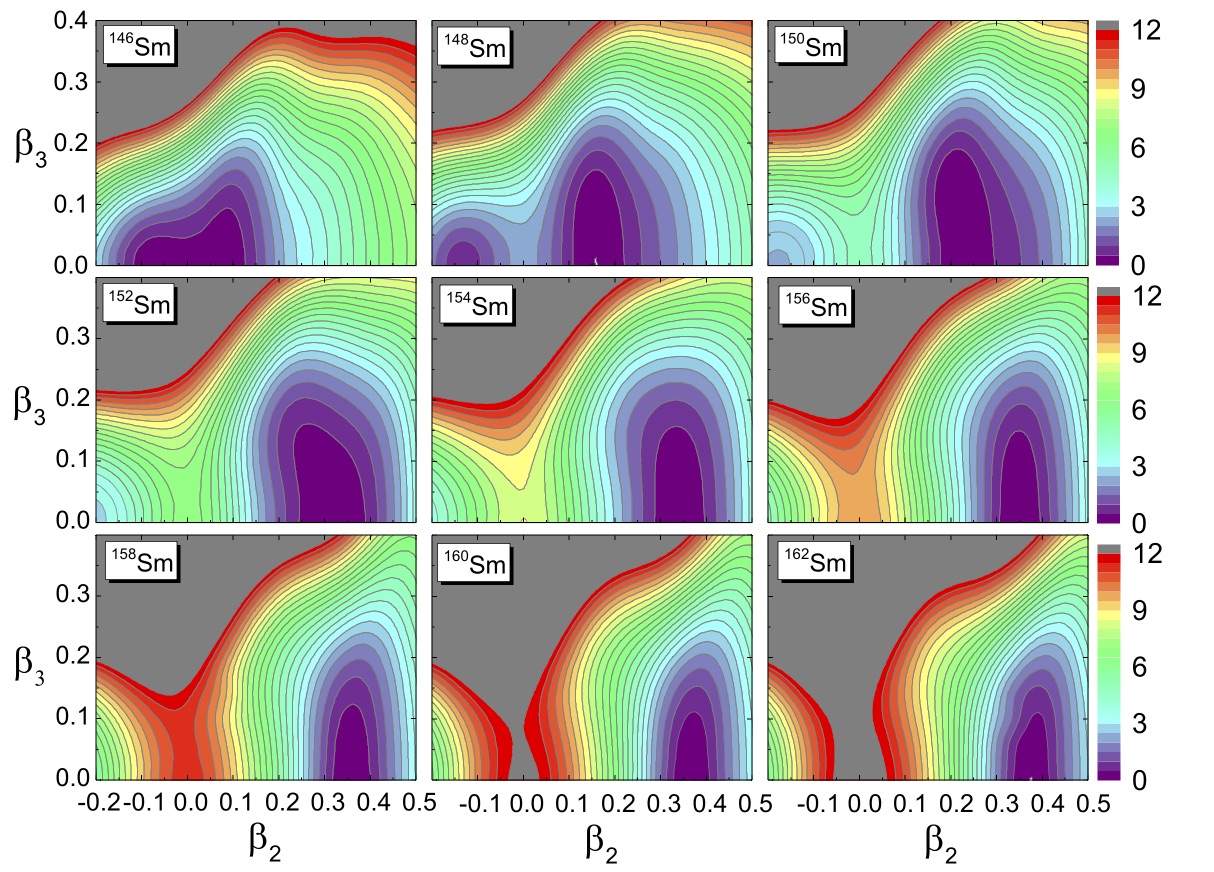}
\includegraphics[height=0.53\textwidth]{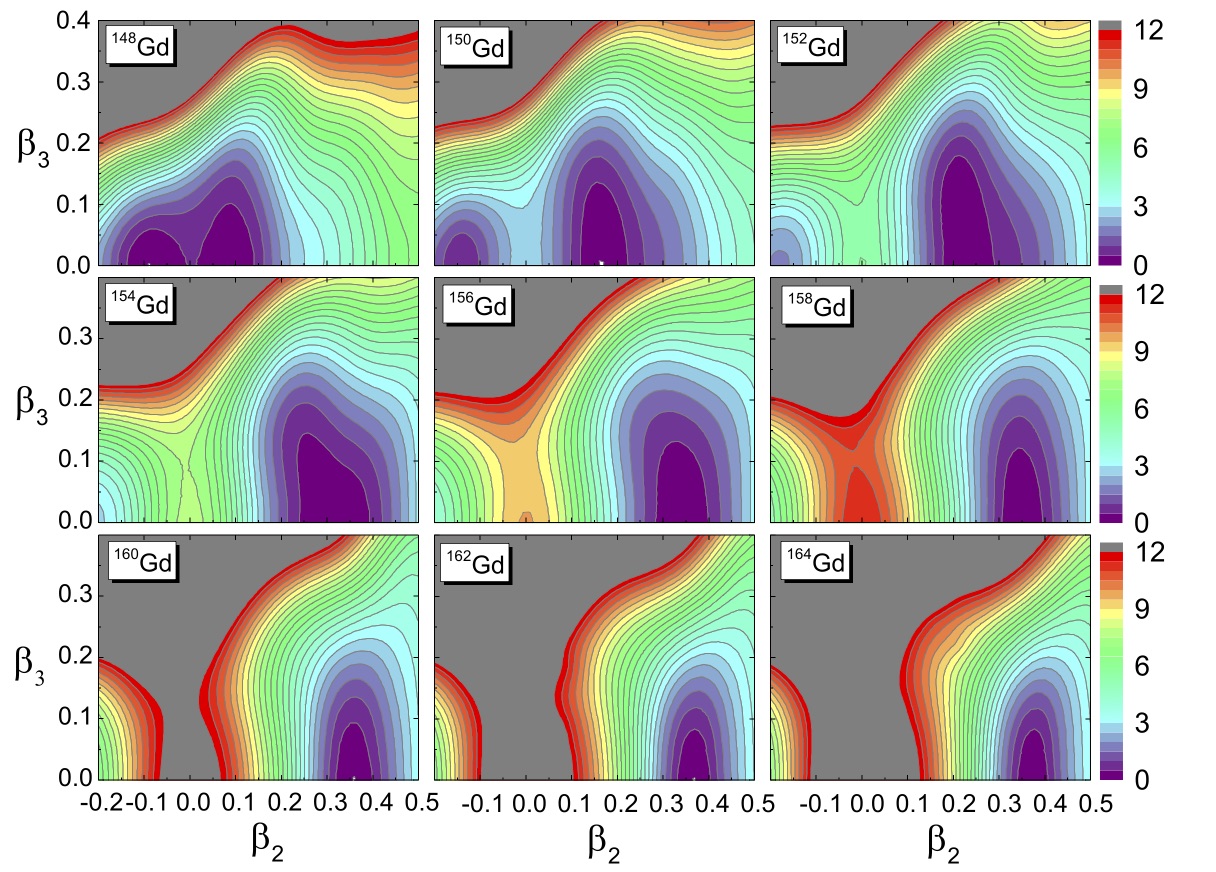}
\caption{(Color online) Same as in the caption to Fig. \ref{PESXeBa} but for the Sm and Gd isotopic chains.}
\label{PESSmGd}
\end{figure}

The principal objective of this study is a systematic analysis that includes collective deformation energy surfaces, excitation energies and average quadrupole and octupole deformations of low-lying states, electric dipole, quadrupole, and octupole transitions for even-even medium-heavy ($54\leq Z\leq 64$ and $84\leq N\leq100$) and heavy ($86\leq Z\leq 100$ and $130\leq N\leq152$) nuclei.

\begin{figure}[ht]
\includegraphics[height=0.39\textwidth]{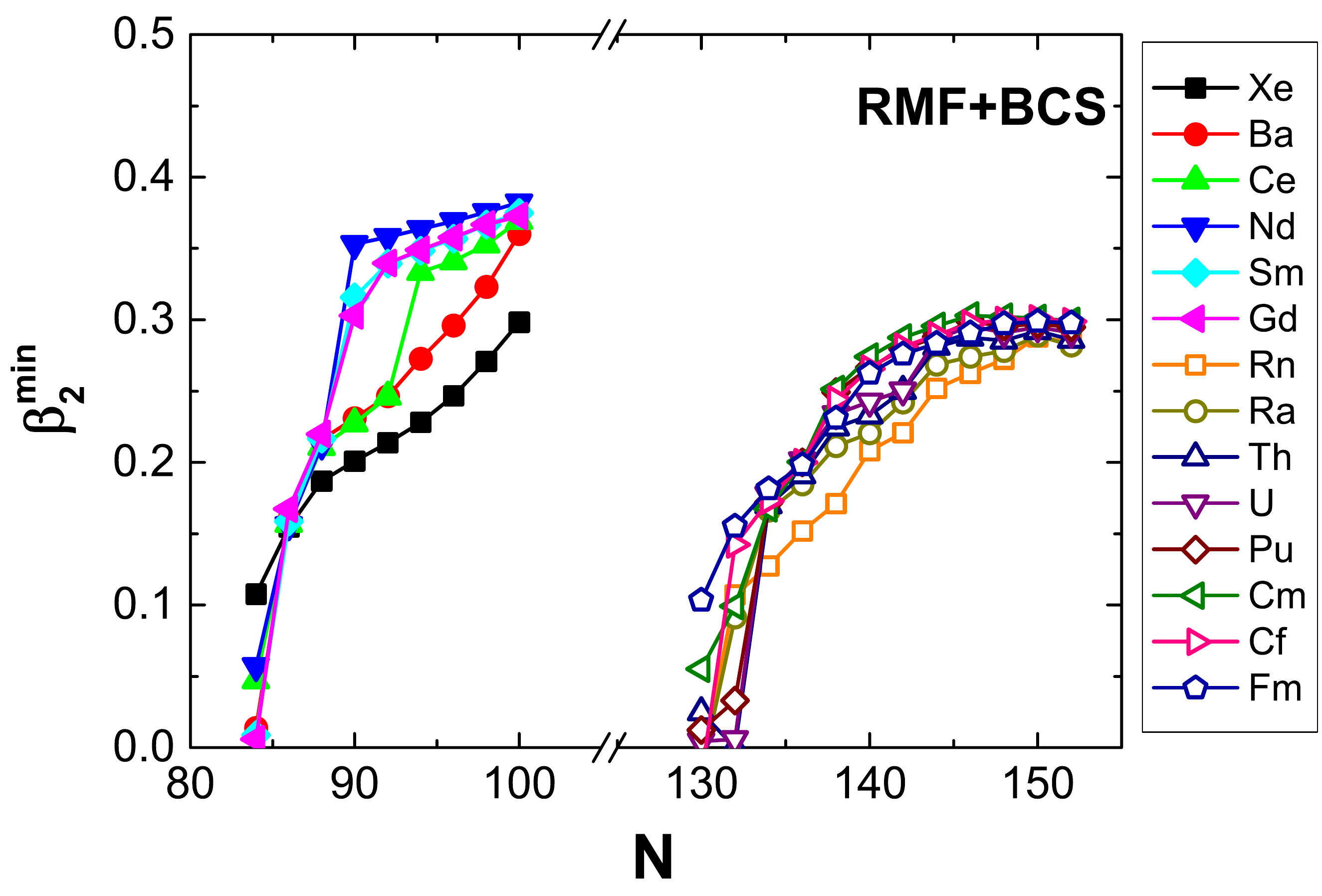}
\includegraphics[height=0.39\textwidth]{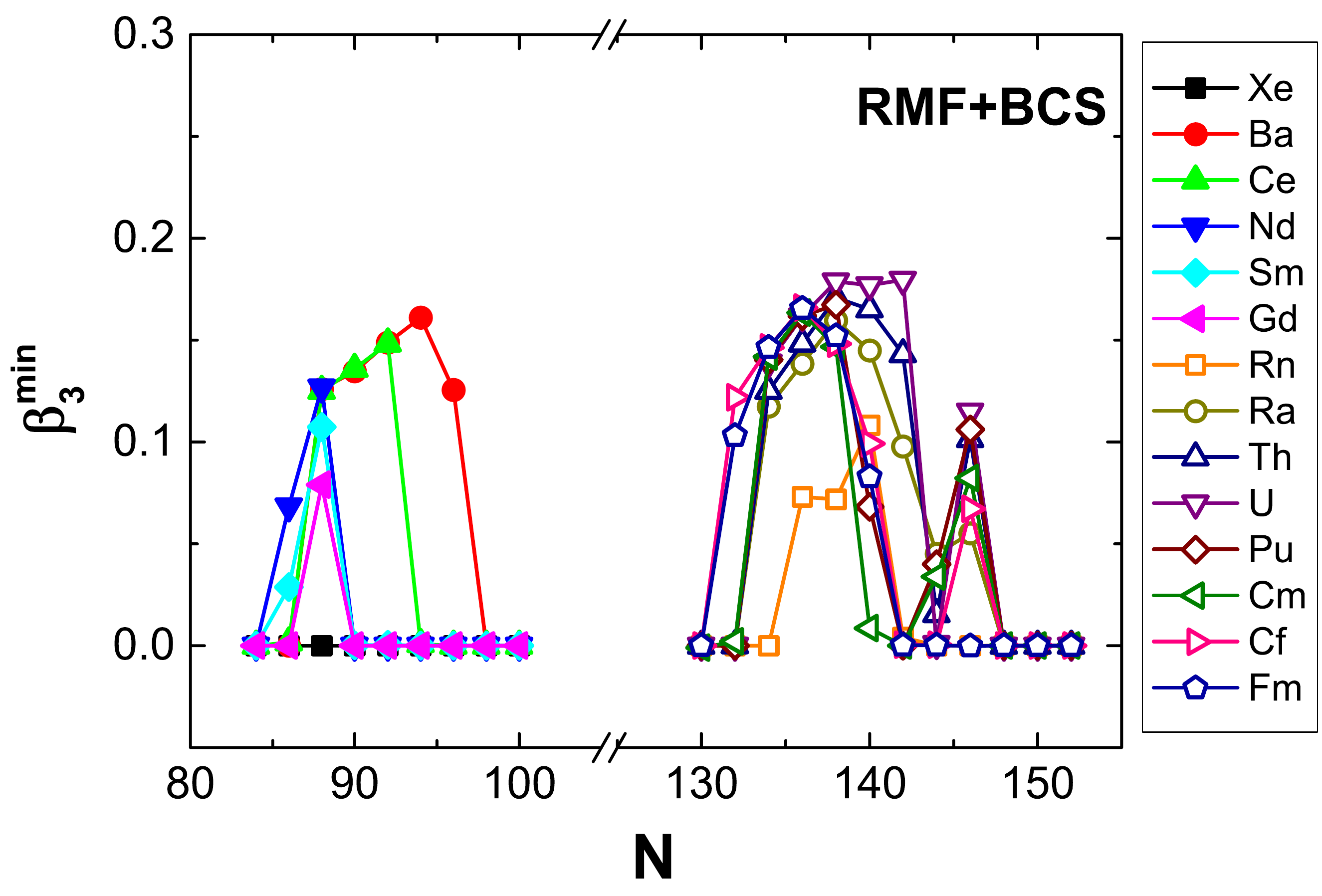}
\caption{(Color online) Calculated values of the equilibrium quadrupole $\beta_2$ and octupole $\beta_3$ deformations as functions of the neutron number,  for the fourteen isotopic chains analyzed in the present study.}
\label{b2b3min}
\end{figure}

Figures \ref{PESXeBa}, \ref{PESCeNd}, and \ref{PESSmGd} display the DESs of the even-even Xe, Ba, Ce, Nd, Sm, and Gd isotopes in the $\beta_2-\beta_3$ plane, calculated with the RMF+BCS model using the functional PC-PK1 \cite{Zhao10} and $\delta$-force pairing with the strength parameters: $V_n (V_p)=353.0 (367.0)$ MeV fm$^3$. The quadrupole and octupole deformations that correspond to the global minima are also plotted in Fig. \ref{b2b3min}. Along the Xe isotopic chain the  equilibrium quadrupole deformation increases gradually, from nearly spherical to well-deformed shapes as the neutron number increases from 84 to 100. The deformation energy surfaces are soft with respect to the octupole degree of freedom for $^{142-148}$Xe, but a non-zero equilibrium octupole deformation is not predicted in these isotopes. Moving to the Ba and Ce isotopes, one finds a similar shape transition as in the Xe isotopic chain but, in addition, a finite value of the equilibrium octupole deformation is predicted for $^{144-152}$Ba and $^{146-150}$Ce. The octupole deformation $\beta_3$ at the equilibrium minimum ranges between 0.13 and 0.16, and the gain in binding caused by the octupole deformation is $\sim 0.5$ MeV. For Nd, Sm, and Gd isotopes the model calculation does not predict stable octupole minima. An interesting result is the soft energy surfaces with respect to both  quadrupole and octupole deformations for the transitional nuclei with $N\sim90$, which can be related to the phenomenon of quantum shape phase transitions. A similar topography of the DES for $^{150}$Sm has been obtained in the HFB calculation using the Gogny D1S and D1M forces \cite{Rodr12}.

\begin{figure}[ht]
\includegraphics[height=0.67\textwidth]{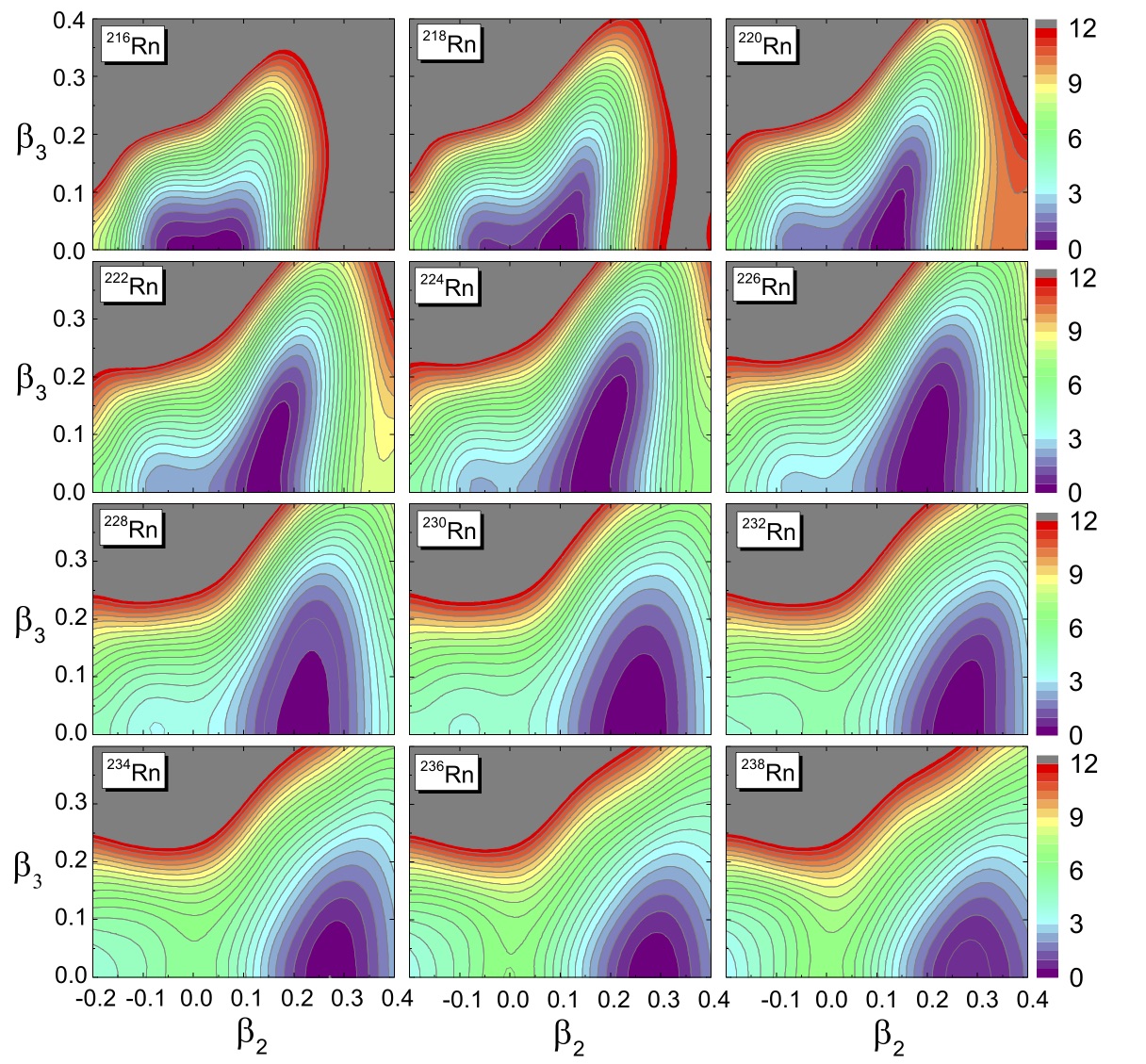}
\includegraphics[height=0.67\textwidth]{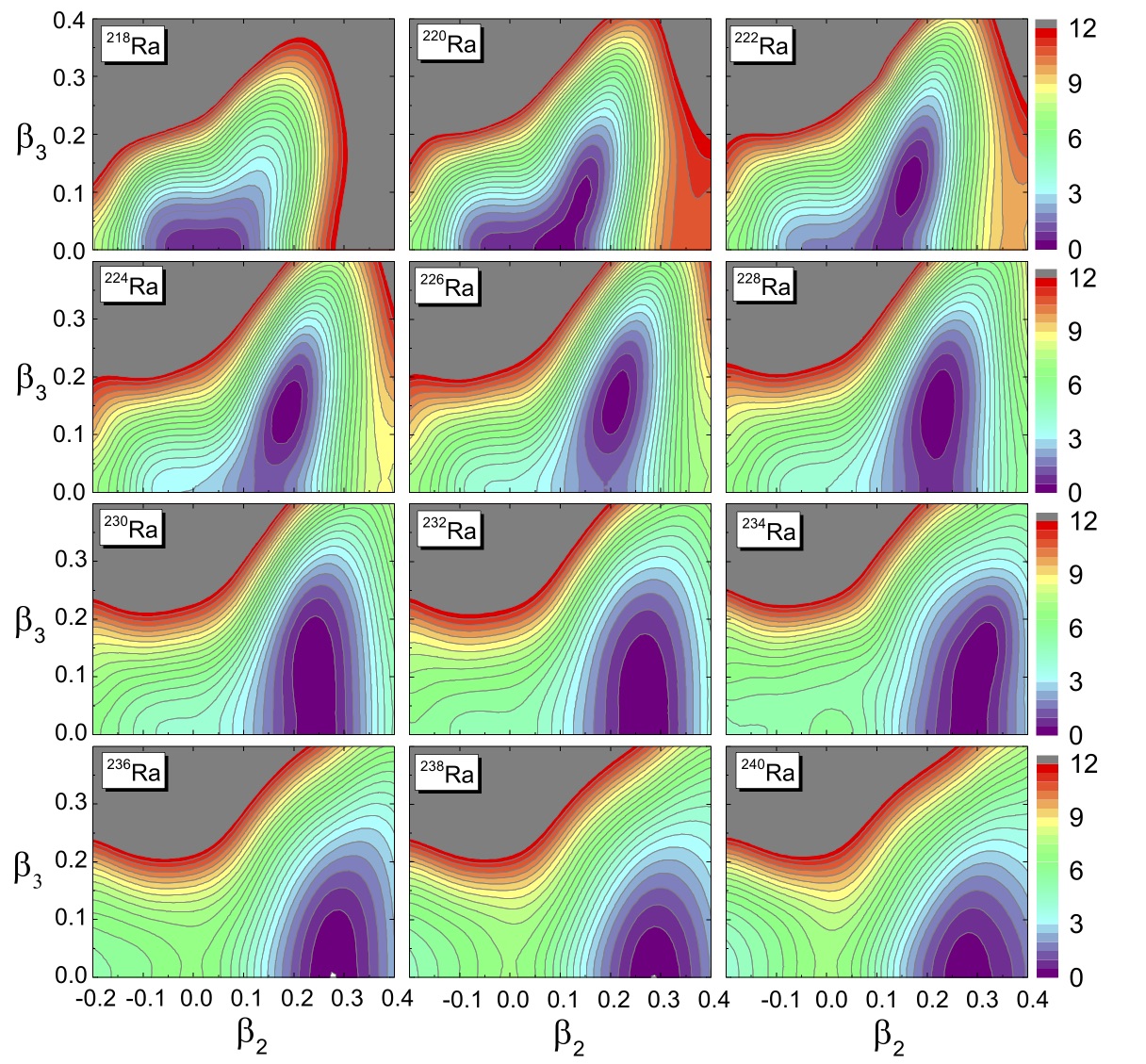}
\caption{(Color online) Deformation energy surfaces of $^{216-238}$Rn and $^{218-240}$Ra in the $\beta_2-\beta_3$ plane, calculated with the RMF+BCS model using the PC-PK1 functional and $\delta$-force pairing. }
\label{PESRn}
\end{figure}

\begin{figure}[ht]
\includegraphics[height=0.67\textwidth]{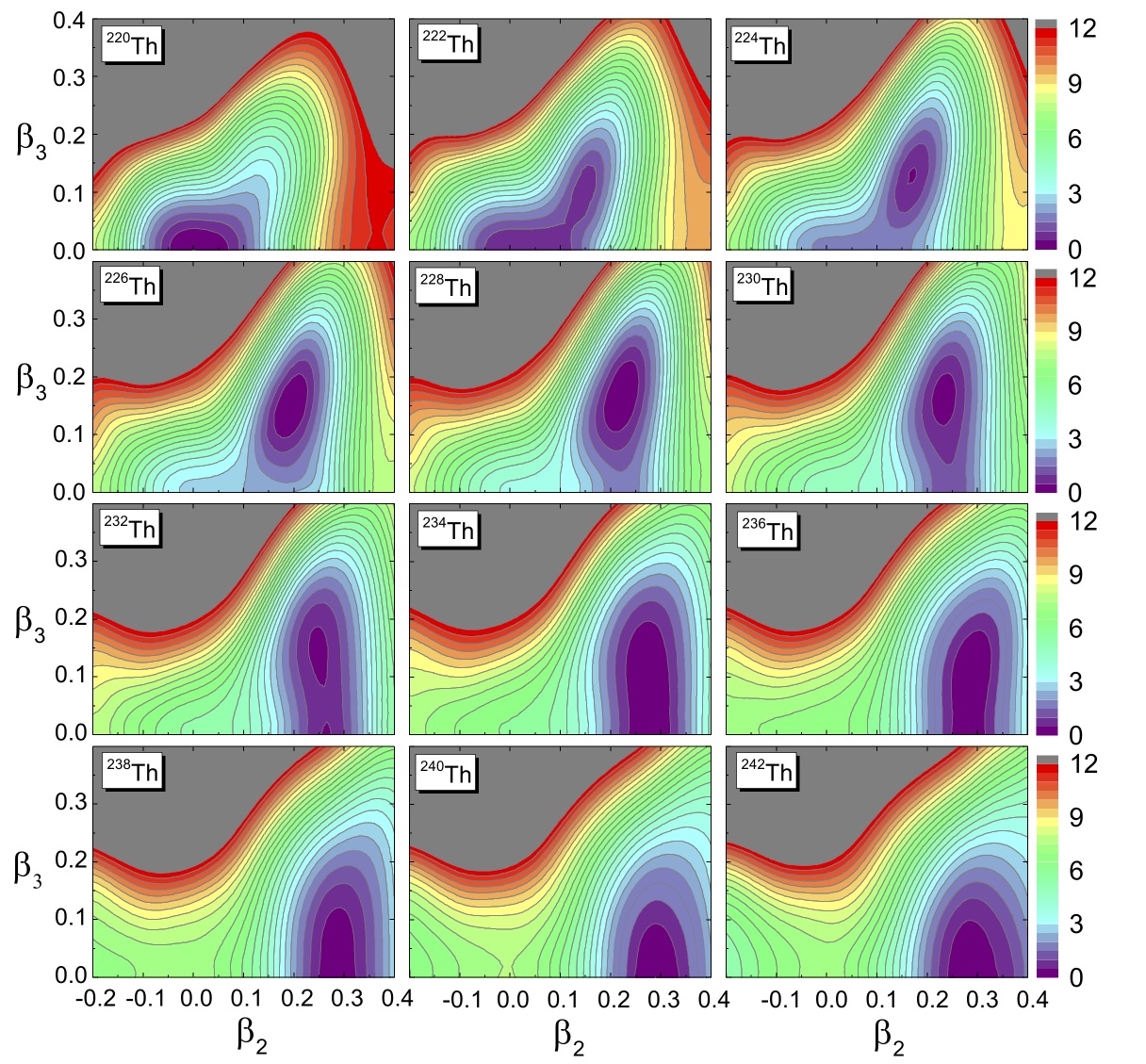}
\includegraphics[height=0.67\textwidth]{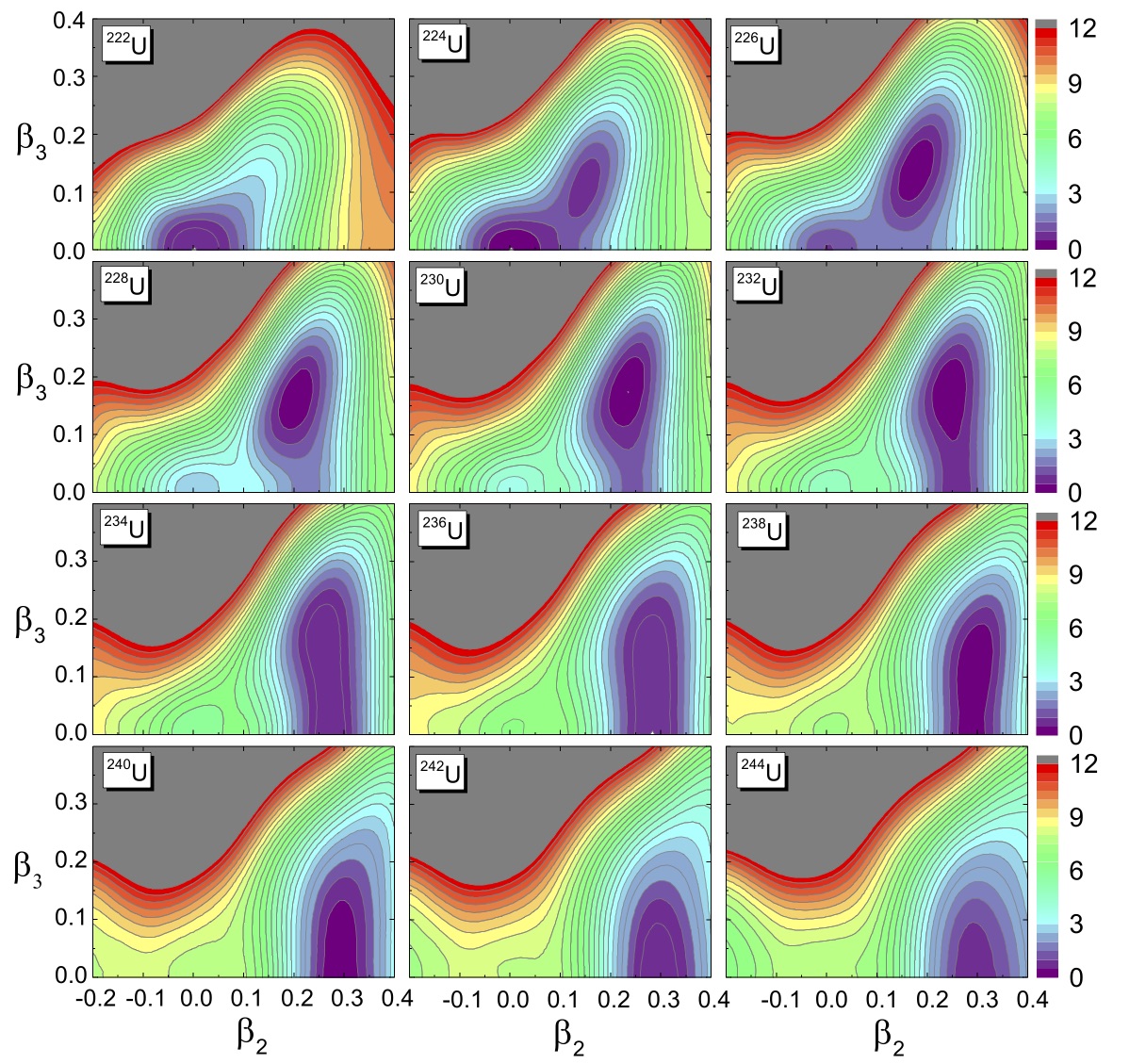}
\caption{(Color online) Same as in the caption to Fig. \ref{PESRn} but for Th and U isotopes.}
\label{PESTh}
\end{figure}

\begin{figure}[ht]
\includegraphics[height=0.65\textwidth]{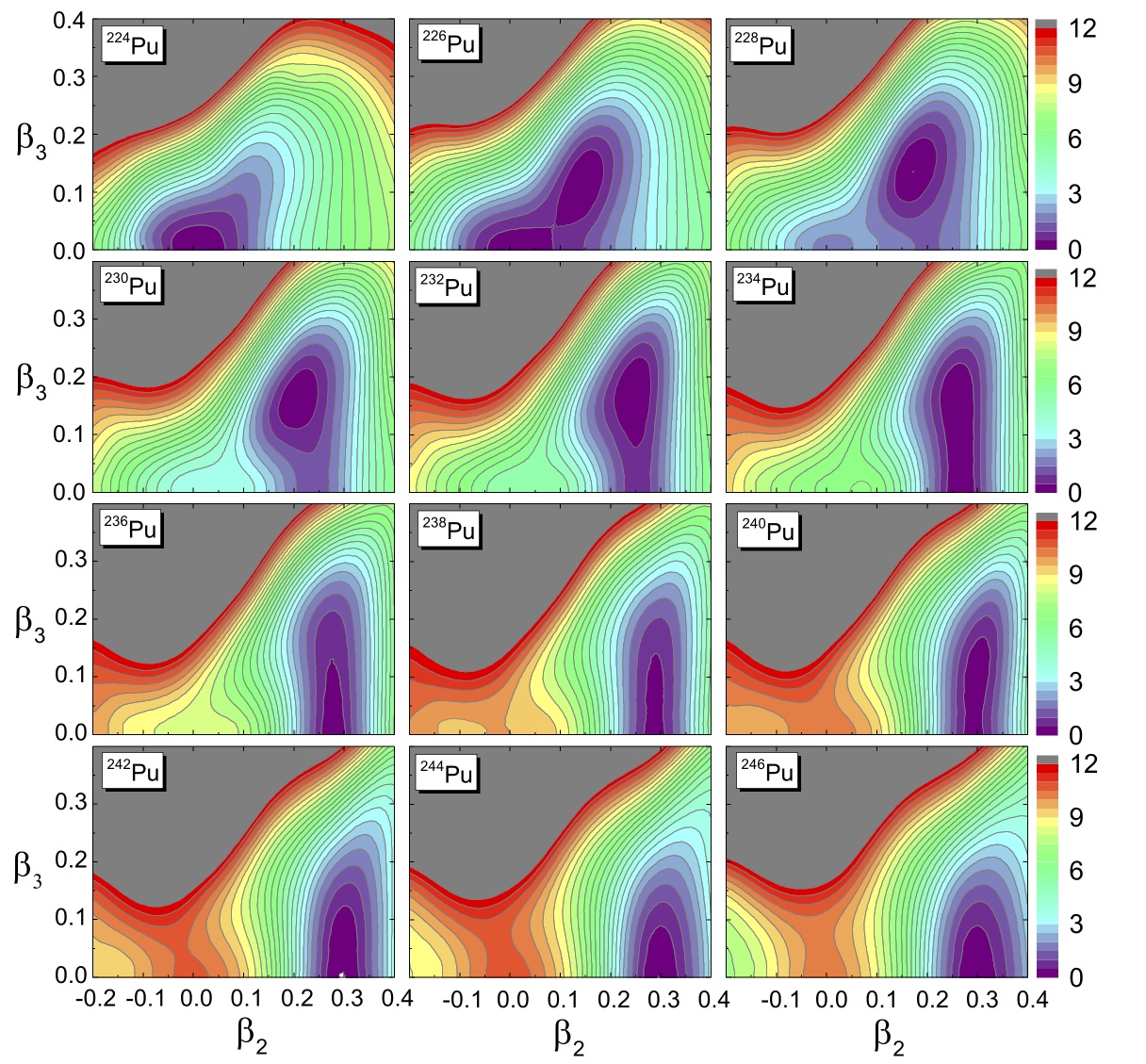}
\includegraphics[height=0.65\textwidth]{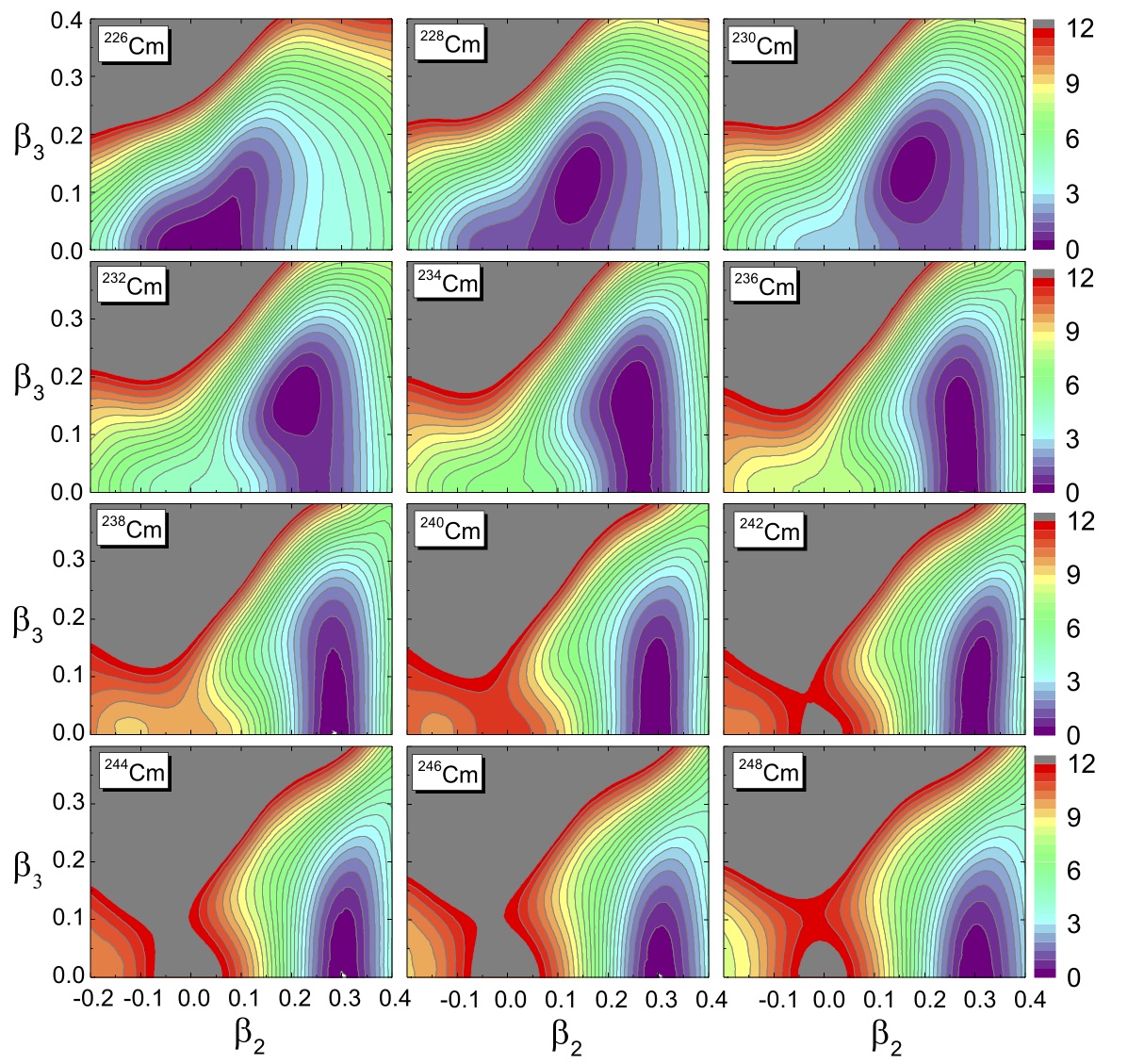}
\caption{(Color online) Same as in the caption to Fig. \ref{PESRn} but for Pu and Cm isotopes.}
\label{PESPu}
\end{figure}

\begin{figure}[ht]
\includegraphics[height=0.65\textwidth]{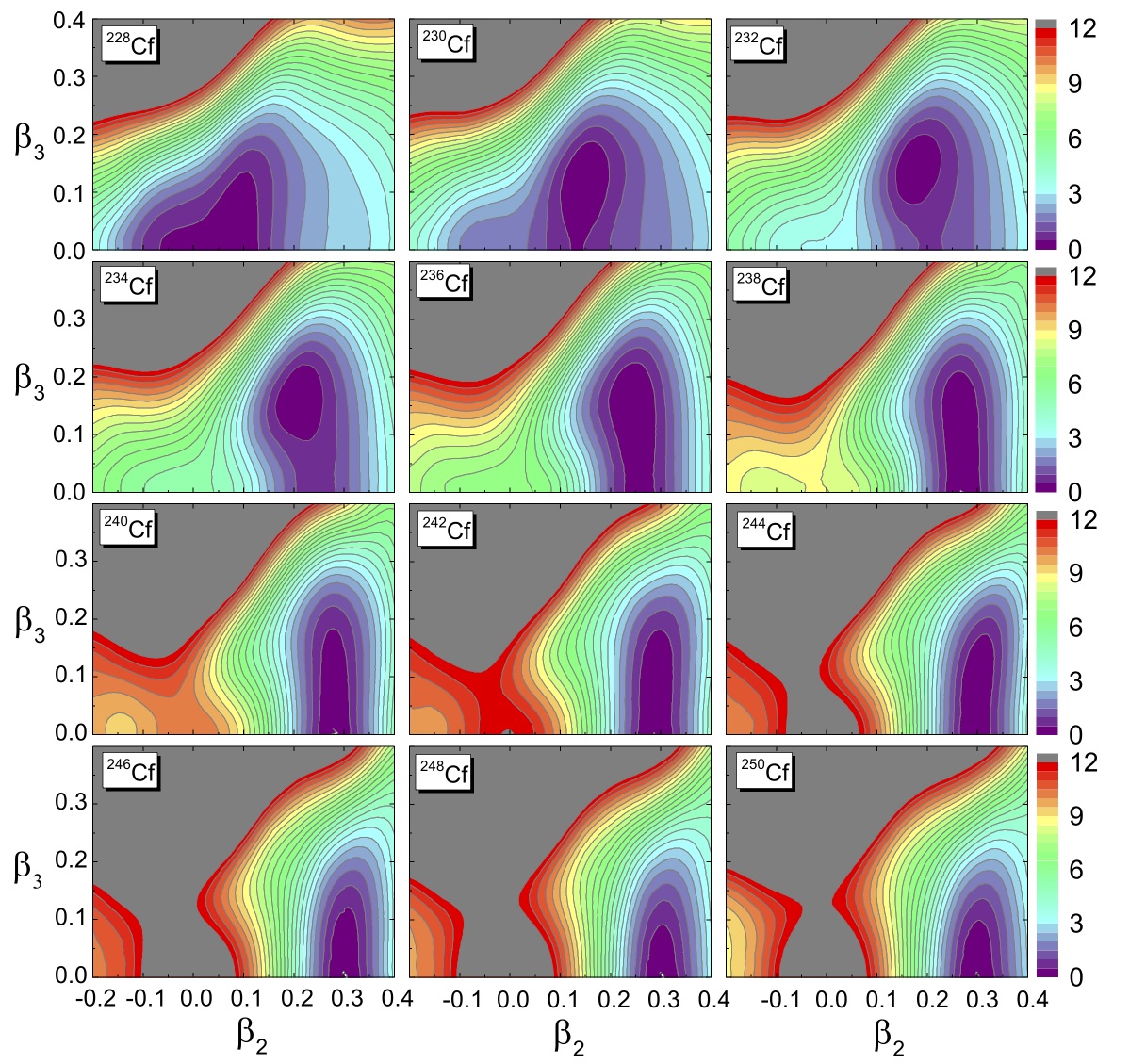}
\includegraphics[height=0.65\textwidth]{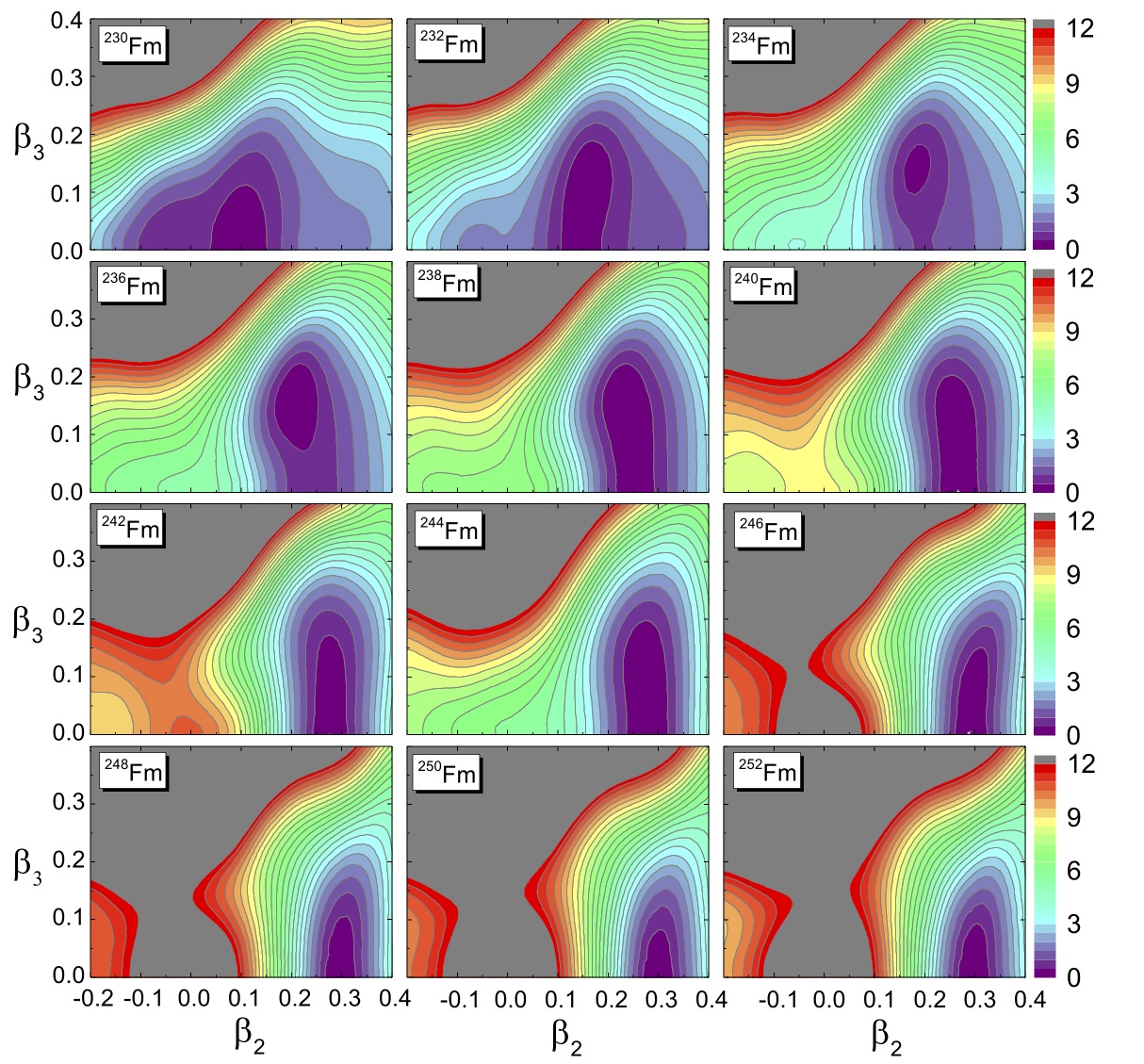}
\caption{(Color online) Same as in the caption to Fig. \ref{PESRn} but for Cf and Fm isotopes.}
\label{PESCf}
\end{figure}

In Figs.~\ref{PESRn}, \ref{PESTh}, \ref{PESPu}, and \ref{PESCf} we plot the deformation energy surfaces of the even-even Rn, Ra, Th, U, Pu, Cm, Cf, and Fm isotopes. In addition, the equilibrium quadrupole and octupole deformation parameters for the eight isotopic chains are shown in Fig. \ref{b2b3min}. All the isotopic chains except Rn exhibit a very interesting shape evolution: from nearly spherical to octupole deformed, octupole soft and, finally, well-deformed prolate quadrupole equilibrium shapes. Stable equilibrium octupole deformations are calculated in $^{222-228}$Ra, $^{224-232}$Th, $^{226-232}$U, $^{228-232}$Pu, $^{228-232}$Cm, $^{230-234}$Cf, and $^{234,236}$Fm. For the Rn isotopic chain, weak octupole deformation is predicted in $^{222-226}$Rn (cf. Fig. \ref{b2b3min}) but the energy surfaces are very shallow with respect to the octupole degree of freedom. Similar shape transitions in the fourteen isotopic chains have also been obtained in studies based on different relativistic energy density functionals \cite{Agbe16,Nomura14}, nonrelativistic functionals \cite{Robledo13,Nomura15}, and macroscopic+microscopic (MM) models \cite{Naza84,Moller08}. Some differences between these calculations are found in the exact location of non-zero equilibrium octupole deformation and the corresponding octupole deformation energies. This can be attributed to the details of the single-particle spectra, especially energy differences between levels with $\Delta j=3$ and $\Delta l=3$, and also to different treatment of pairing correlations \cite{Agbe16}.

\clearpage

\begin{figure}[ht]
\includegraphics[height=0.4\textwidth]{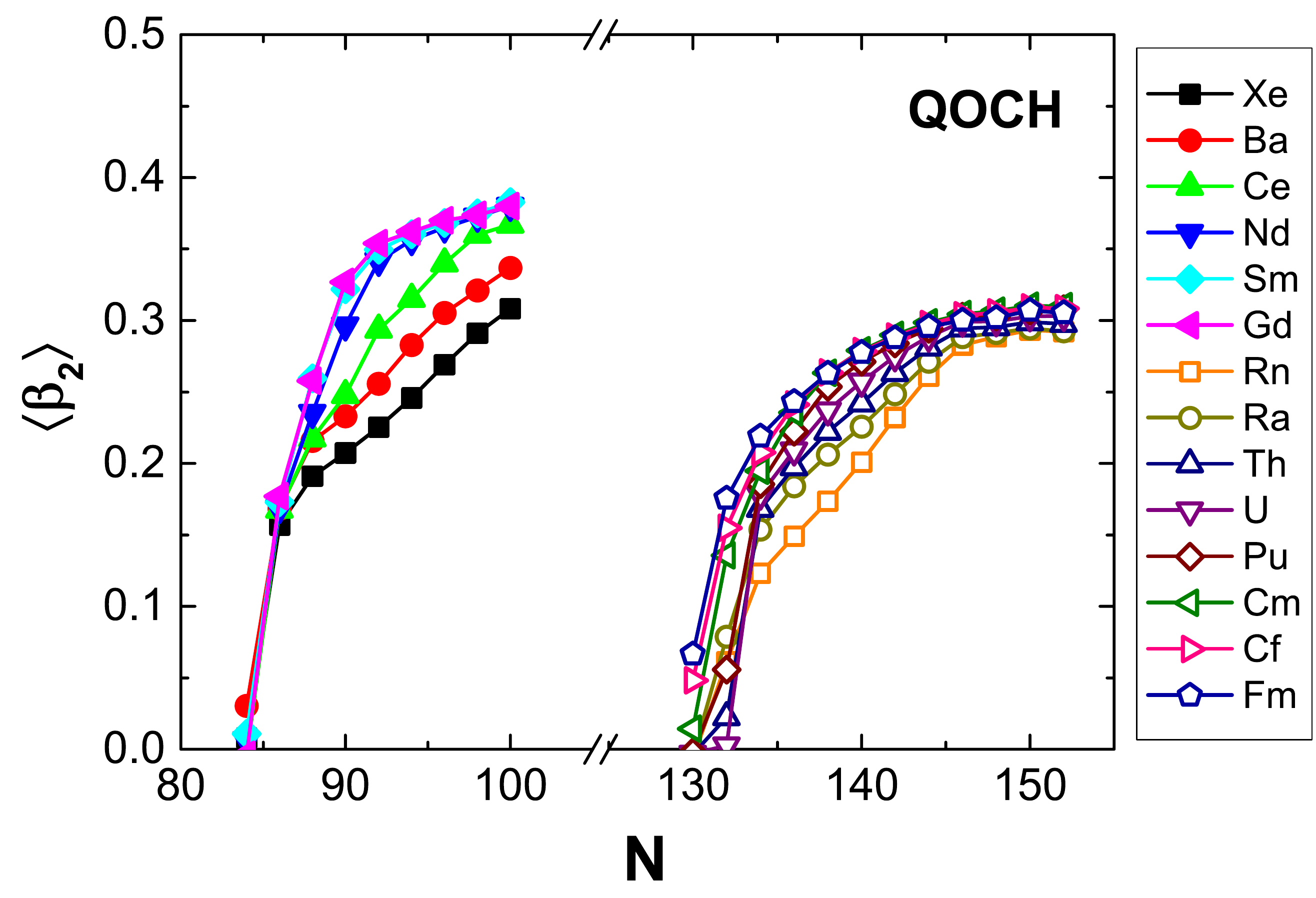}
\includegraphics[height=0.4\textwidth]{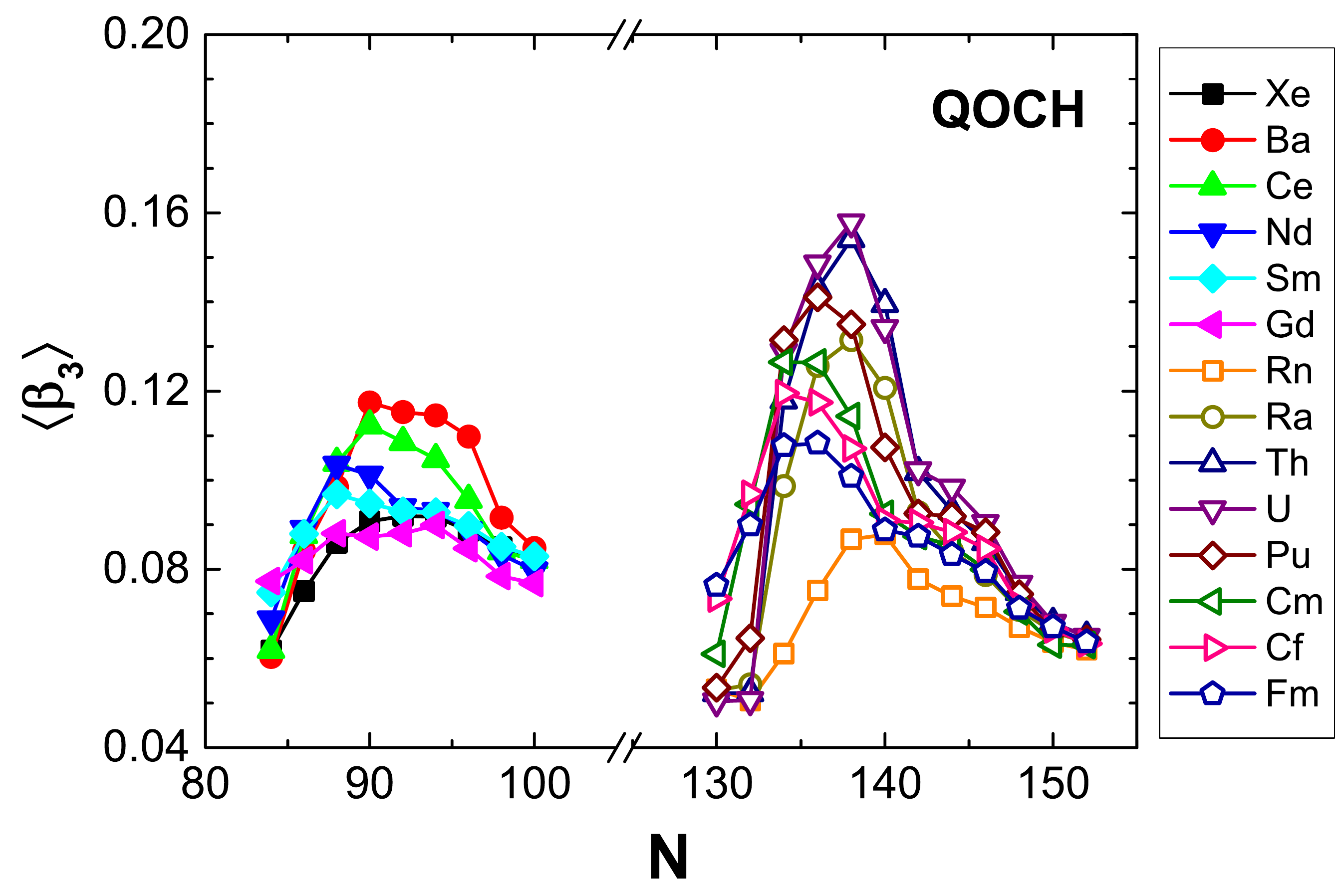}
\includegraphics[height=0.4\textwidth]{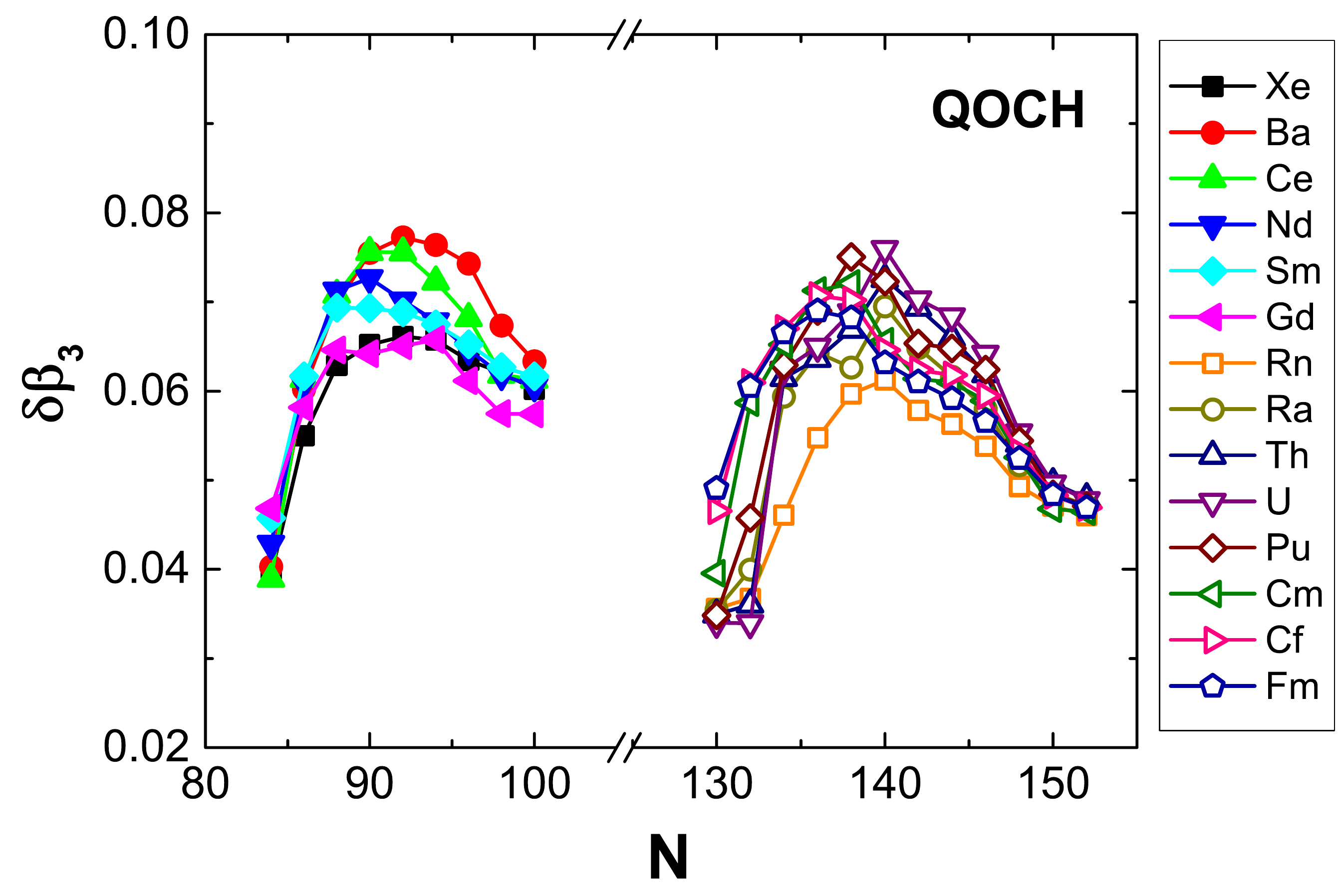}
\caption{(Color online) Mean values of the quadrupole $\langle\beta_2\rangle$ and octupole $\langle\beta_3\rangle$ deformations, as well as the octupole variance $\delta\beta_3=\sqrt{\langle\beta_3^2\rangle-\langle\beta_3\rangle^2}$, computed for the QOCH ground states $0^+_1$, as functions of the neutron number.}
\label{aveb2b3}
\end{figure}

Figure \ref{aveb2b3} displays the expectation values of the quadrupole $\langle\beta_2\rangle$ and octupole $\langle\beta_3\rangle$ deformations, as well as the octupole variance $\delta\beta_3=\sqrt{\langle\beta_3^2\rangle-\langle\beta_3\rangle^2}$, in the QOCH ground states $0^+_1$, as functions of the neutron number. Initially the ground-state quadrupole deformation $\langle\beta_2\rangle$ increases rapidly and then more gradually with neutron number, in both mass regions.  The corresponding ground-state octupole deformation $\langle\beta_3\rangle$ increases at first, and then decreases with peaks at $N\sim90$ for medium-heavy nuclei, and at $N\sim136$ for heavy nuclei. In our calculation $\langle\beta_3\rangle\gtrsim0.12$ is predicted for octupole deformed nuclei, whereas $\langle\beta_3\rangle\sim0.09$ for nuclei with octupole soft DESs. Both octupole deformed and octupole soft nuclei exhibit large shape fluctuations, quantified by the variance $\delta\beta_3=\sqrt{\langle\beta_3^2\rangle-\langle\beta_3\rangle^2}$ that is shown in the bottom panel of Fig.~\ref{aveb2b3}.

\clearpage
\begin{figure}[ht]
\includegraphics[height=0.8\textwidth]{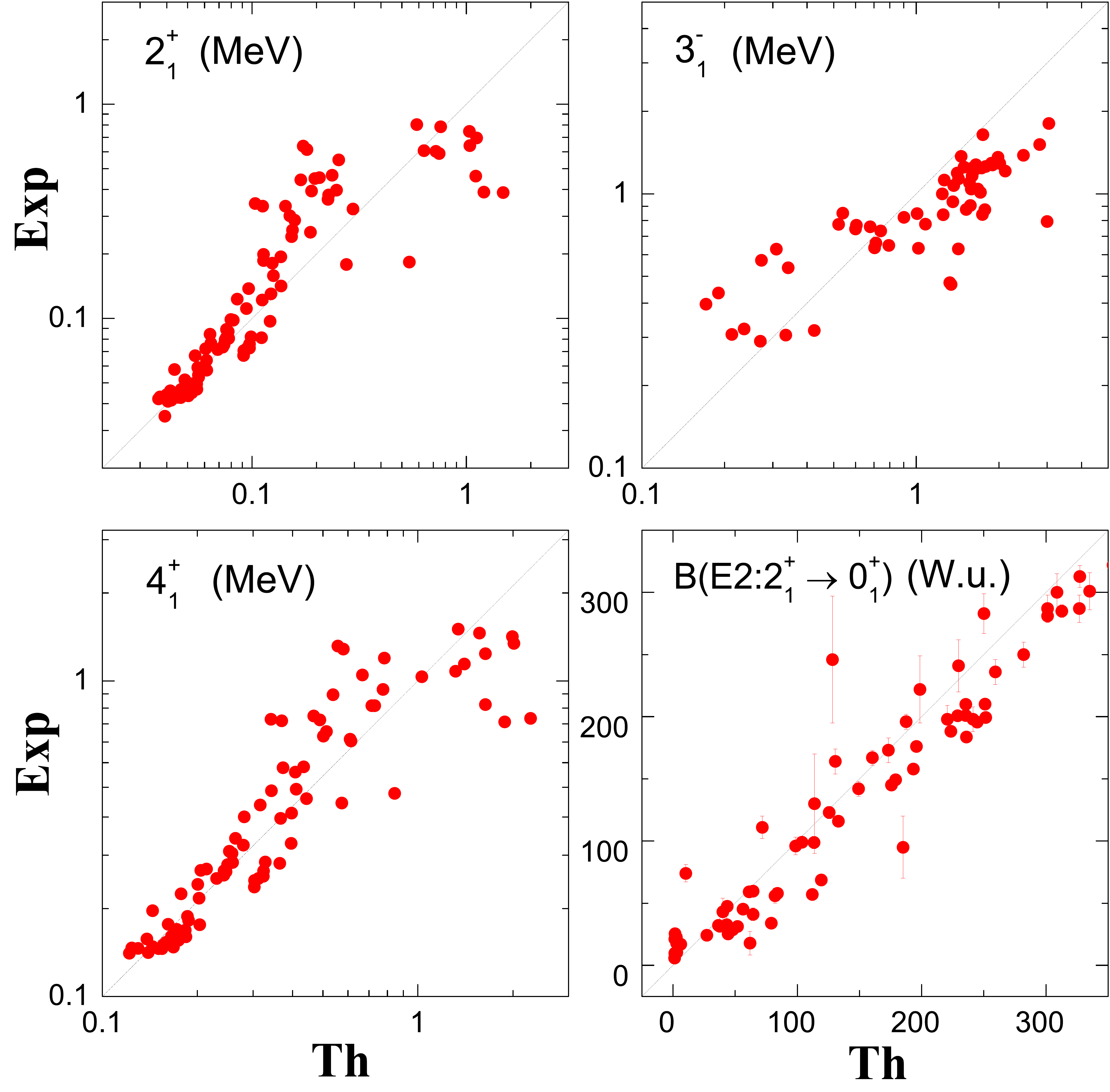}
\caption{(Color online) Comparison between the theoretical excitation energies of the states $2^+_1$, $4^+_1$, and $3^-_1$, and the $B(E2; 2^+_1\to0^+_1)$ values, and their experimental counterparts \cite{NNDC}.}
\label{ExpTh}
\end{figure}

\begin{figure}[ht]
\includegraphics[height=0.384\textwidth]{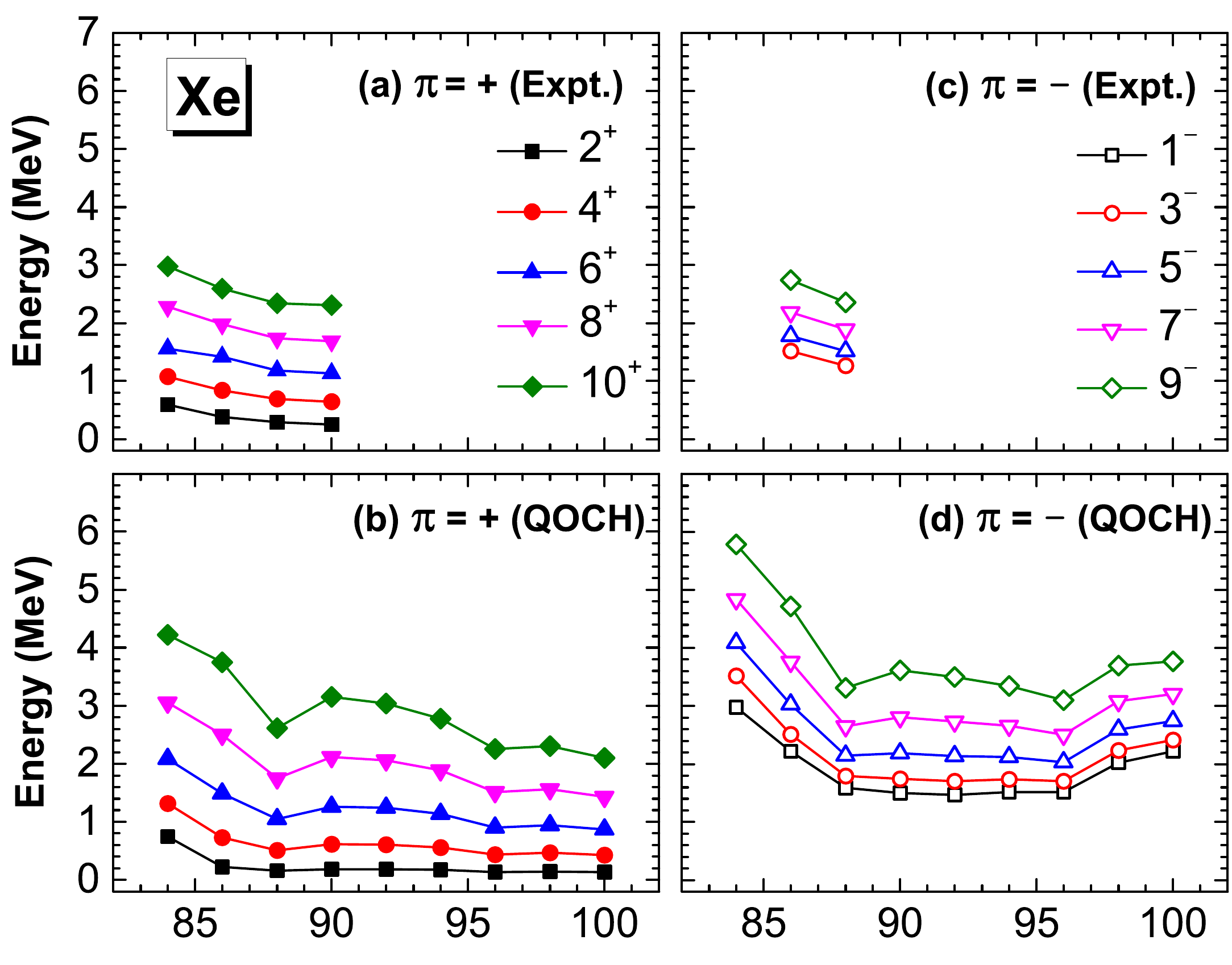}
\includegraphics[height=0.384\textwidth]{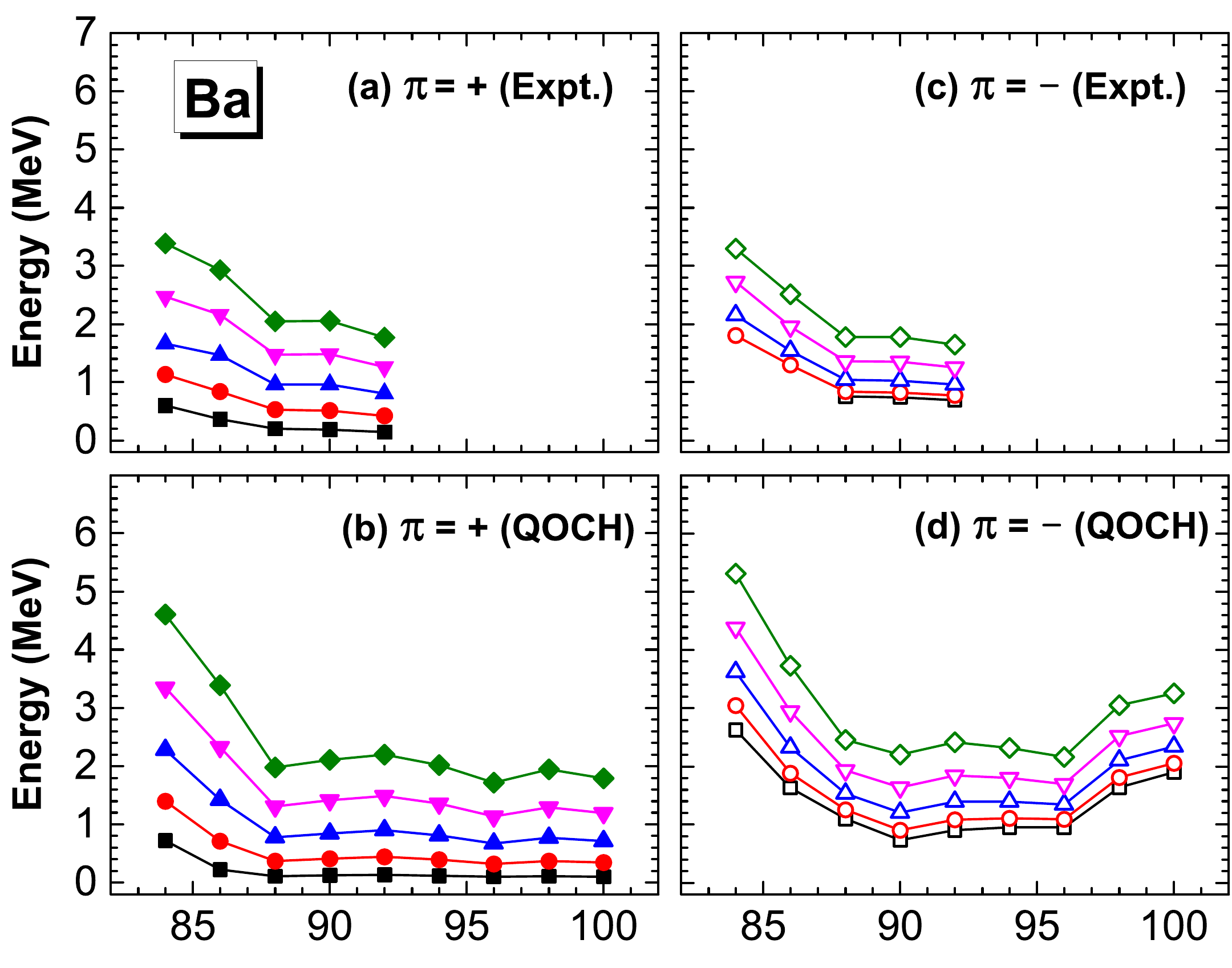}
\includegraphics[height=0.384\textwidth]{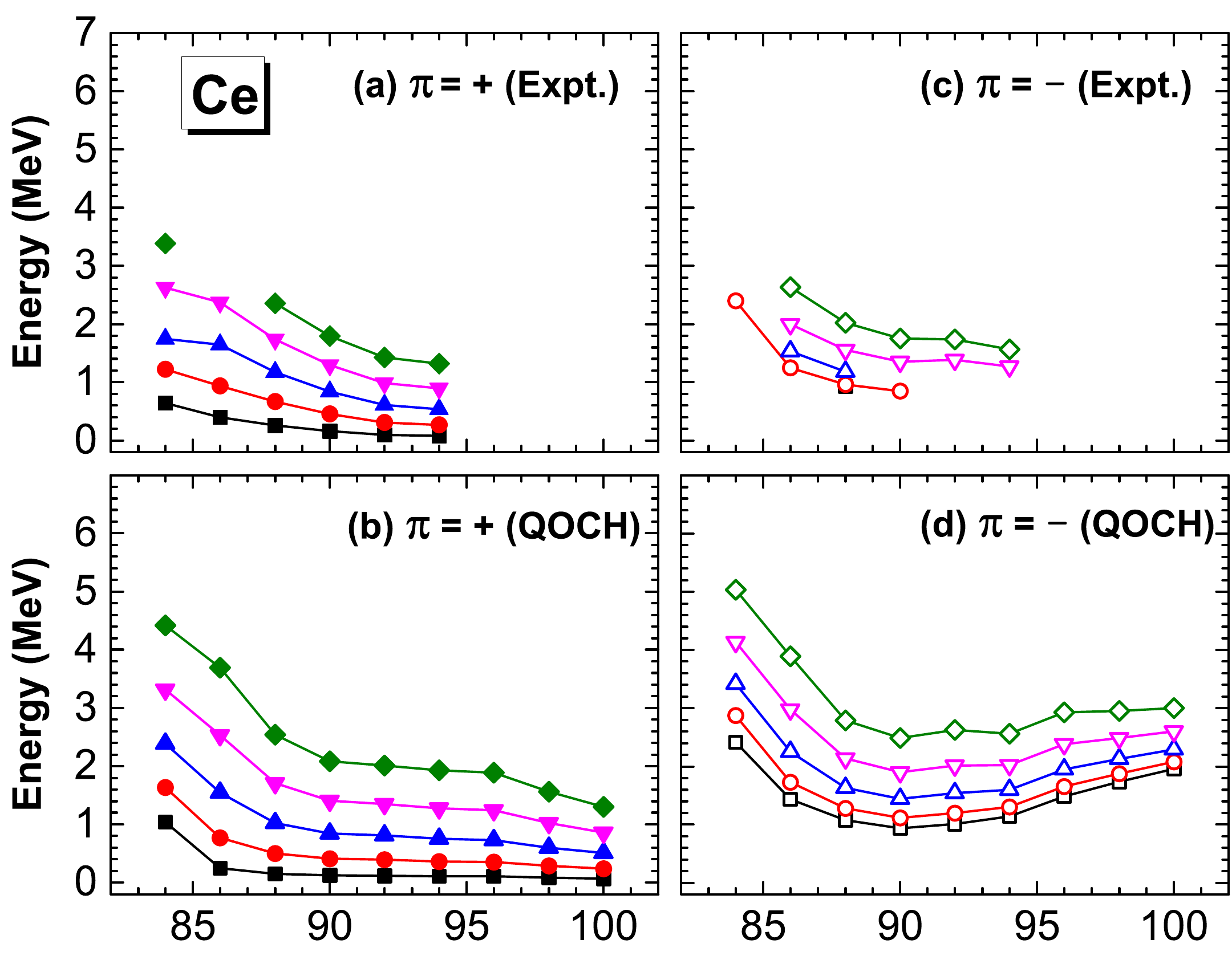}
\includegraphics[height=0.384\textwidth]{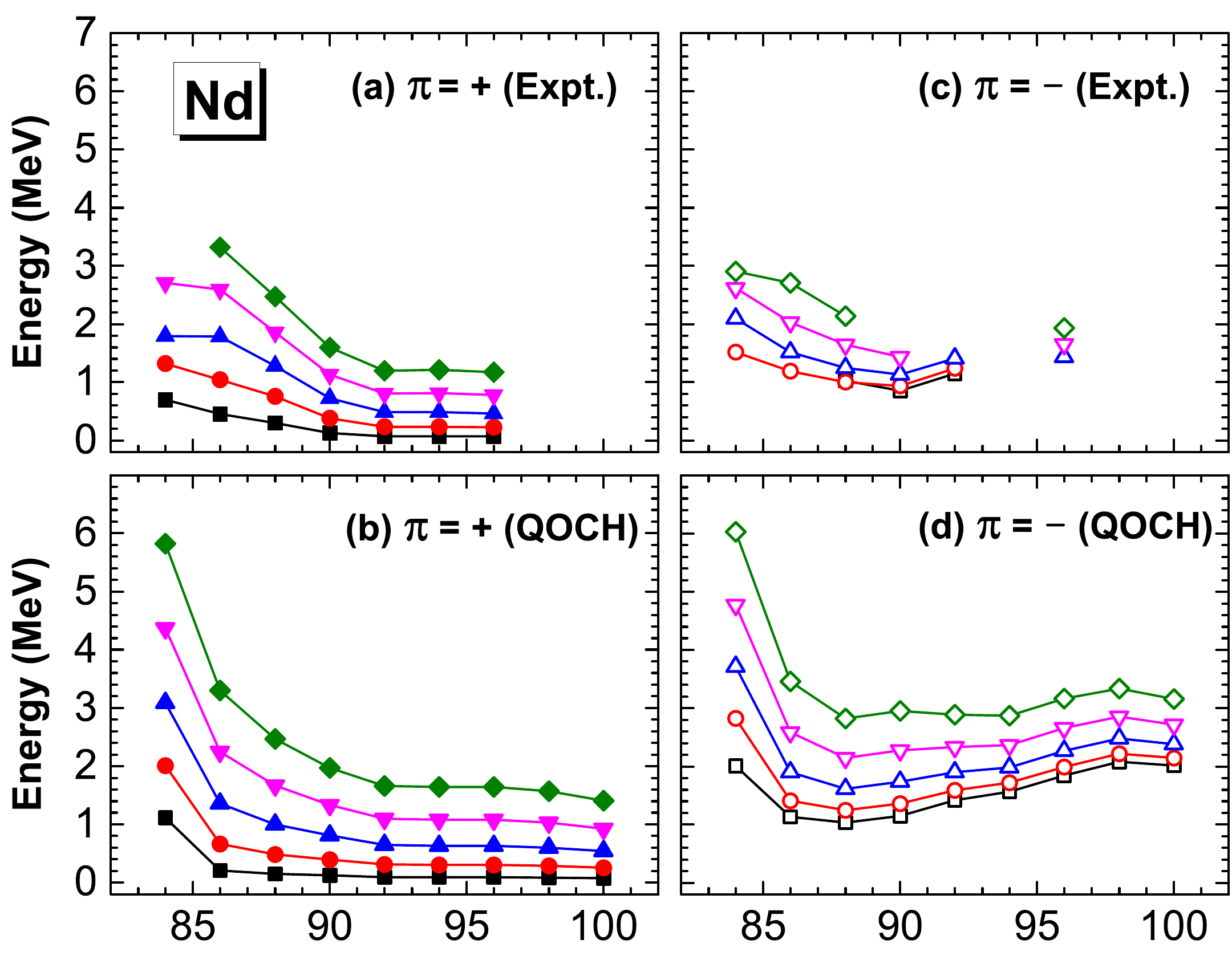}
\includegraphics[height=0.4\textwidth]{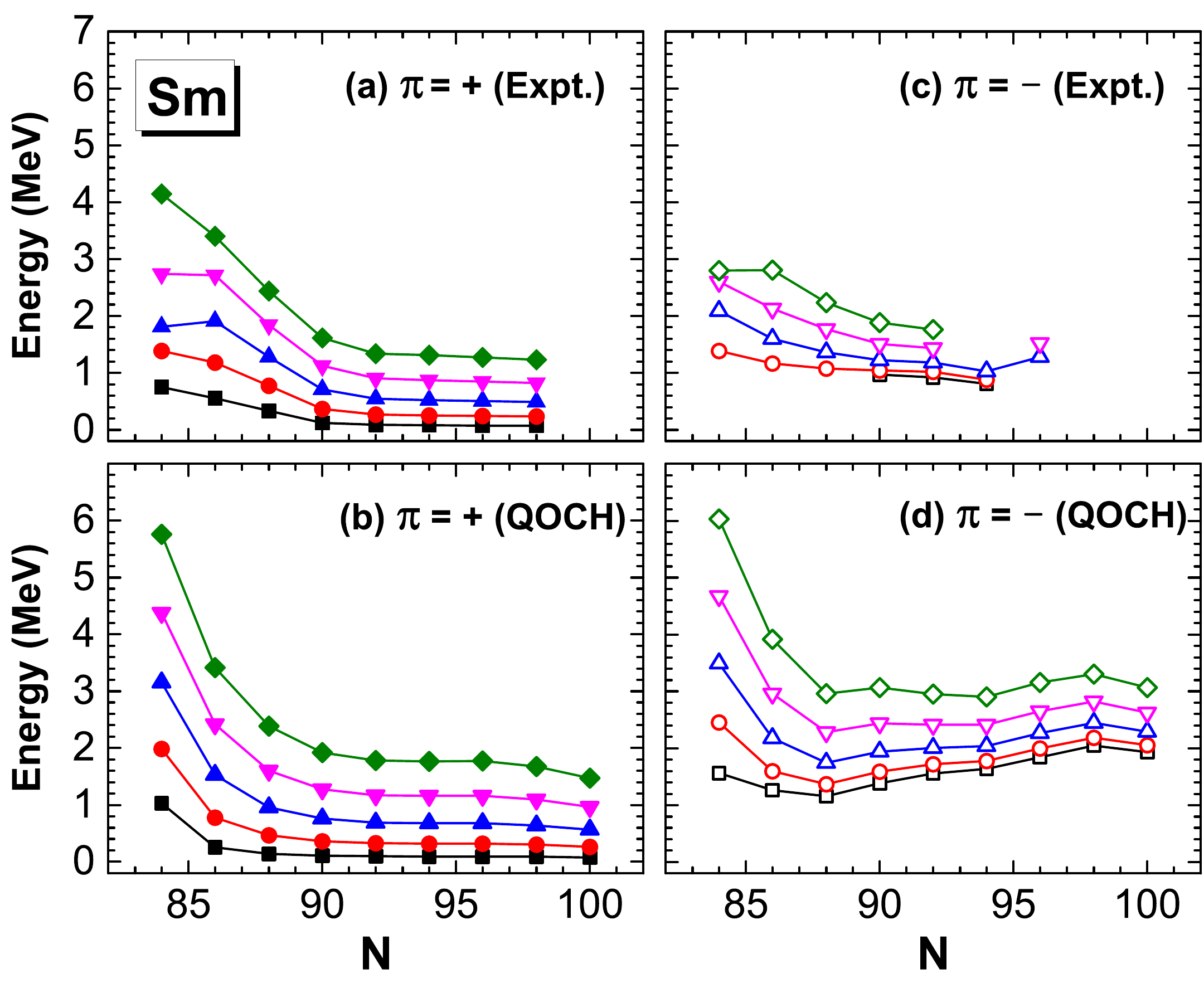}
\includegraphics[height=0.4\textwidth]{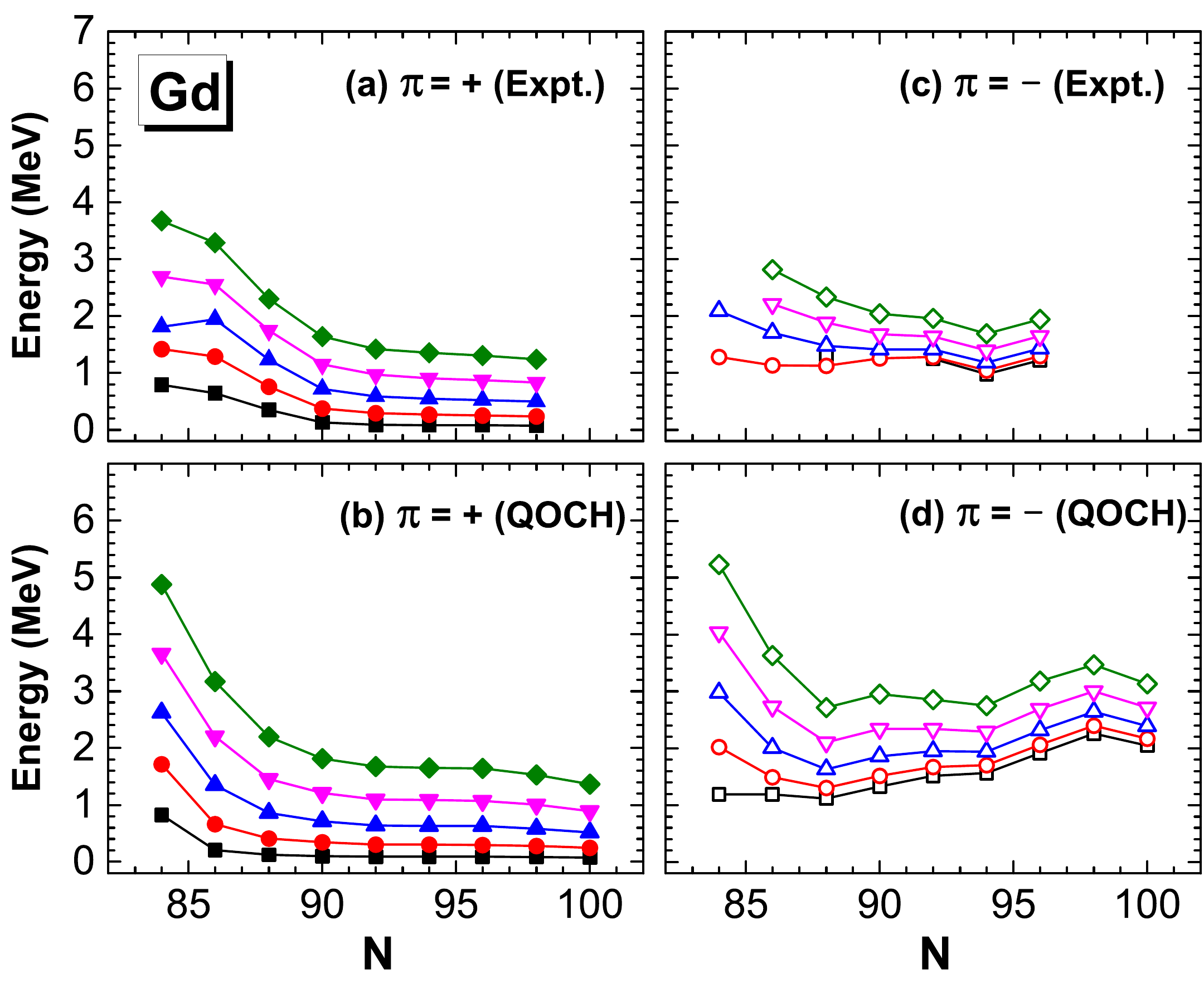}
\caption{(Color online) The energy spectra of the low-lying even-spin positive-parity states up to $J^\pi=10^+$ (panels a,b), and odd-spin negative-parity states up to $J^\pi=9^-$ (panels c,d), as functions of the neutron number for the Xe, Ba, Ce, Nd, Sm, and Gd isotopes. The experimental values are from the NNDC compilation \cite{NNDC}.}
\label{specLa}
\end{figure}

\begin{figure}[ht]
\includegraphics[height=0.37\textwidth]{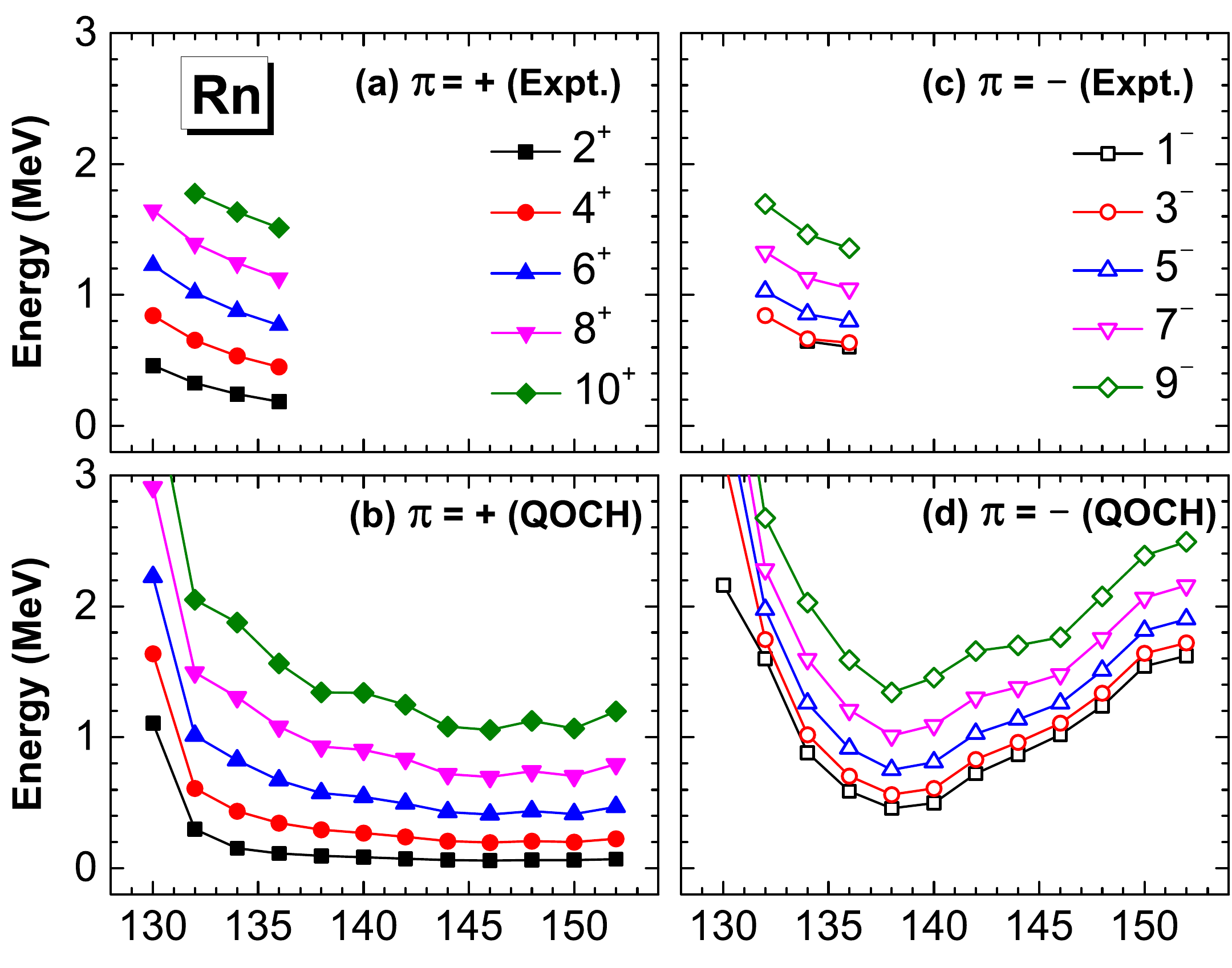}
\includegraphics[height=0.37\textwidth]{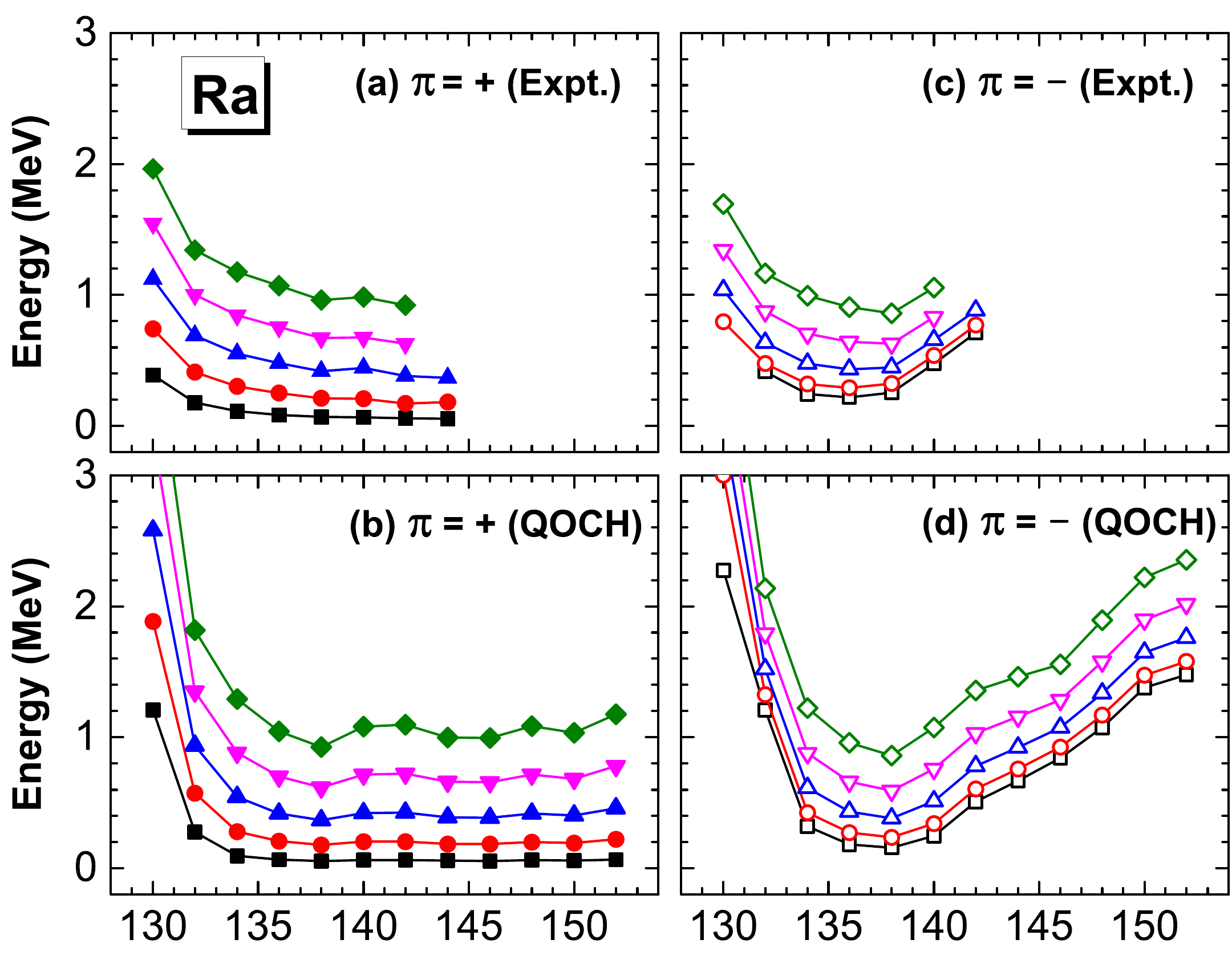}\vspace{-0.1cm}
\includegraphics[height=0.37\textwidth]{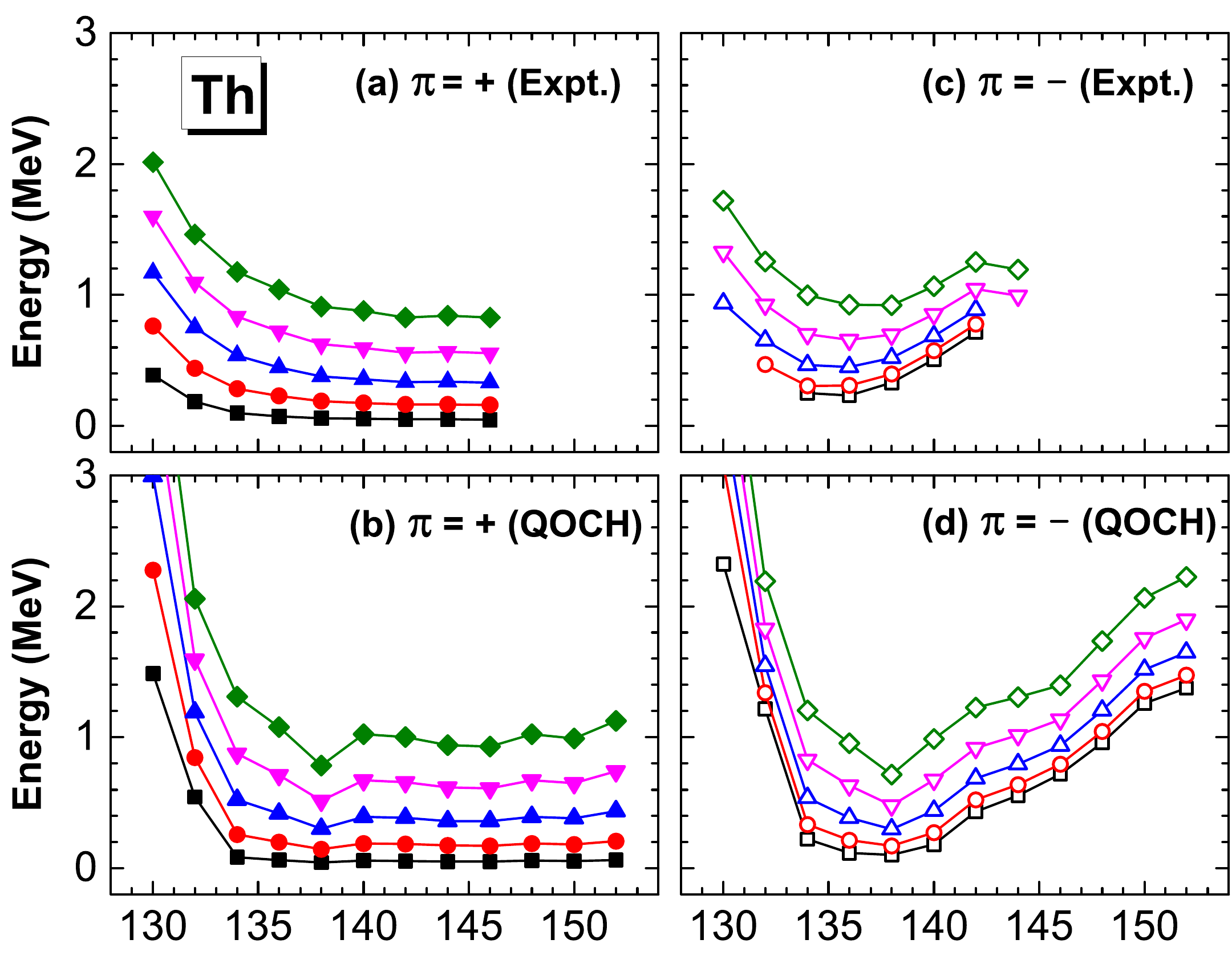}
\includegraphics[height=0.37\textwidth]{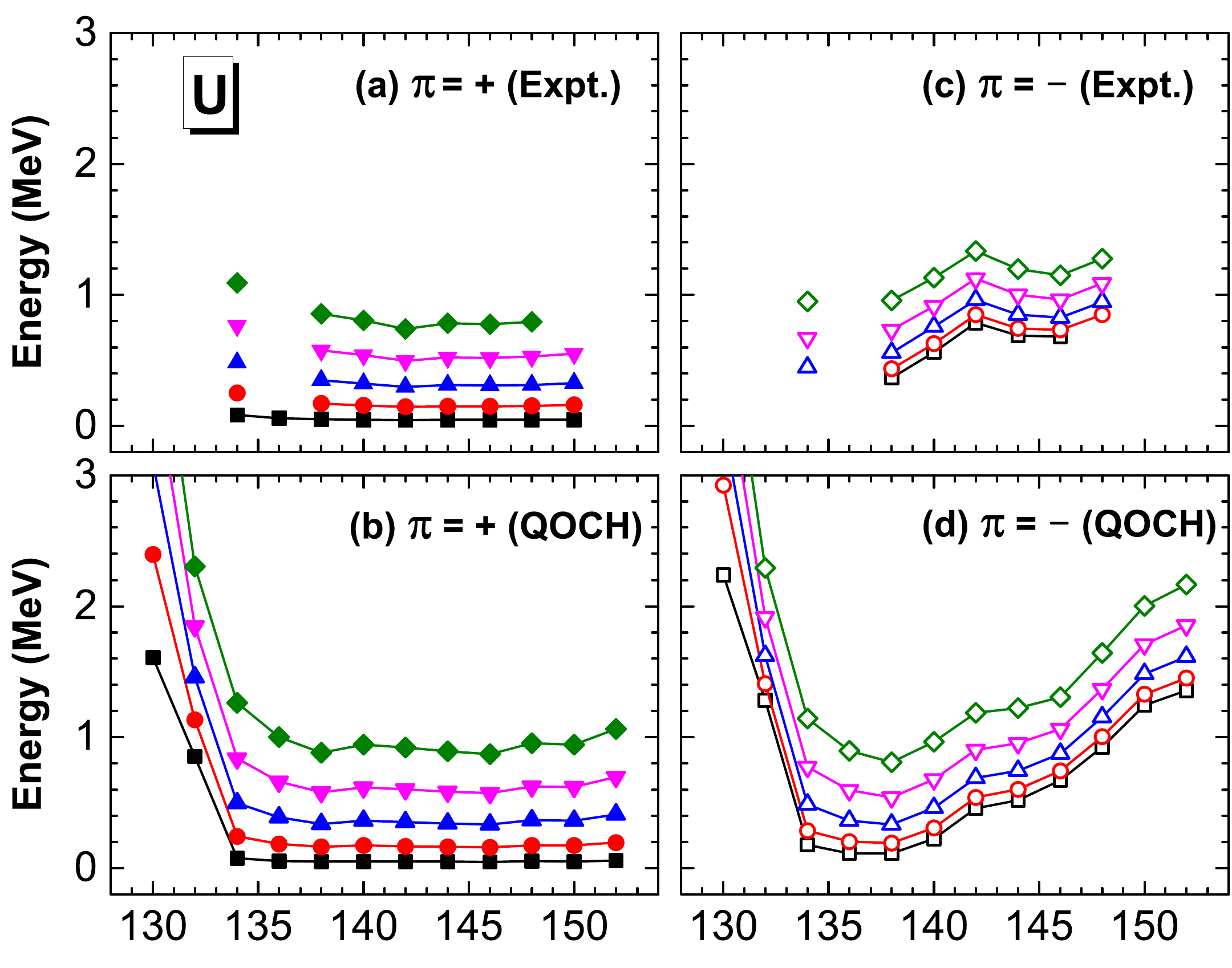}\vspace{-0.1cm}
\includegraphics[height=0.37\textwidth]{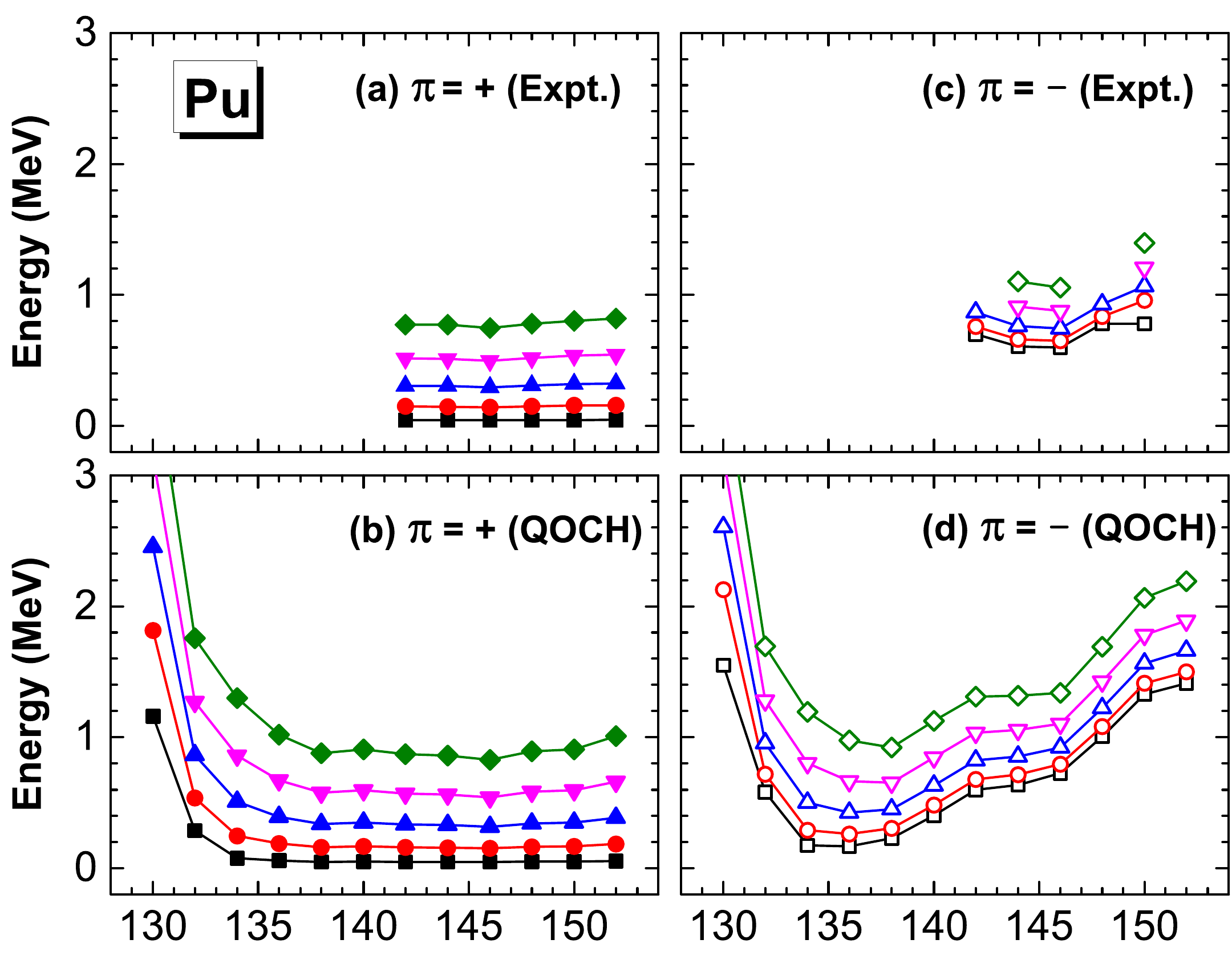}
\includegraphics[height=0.37\textwidth]{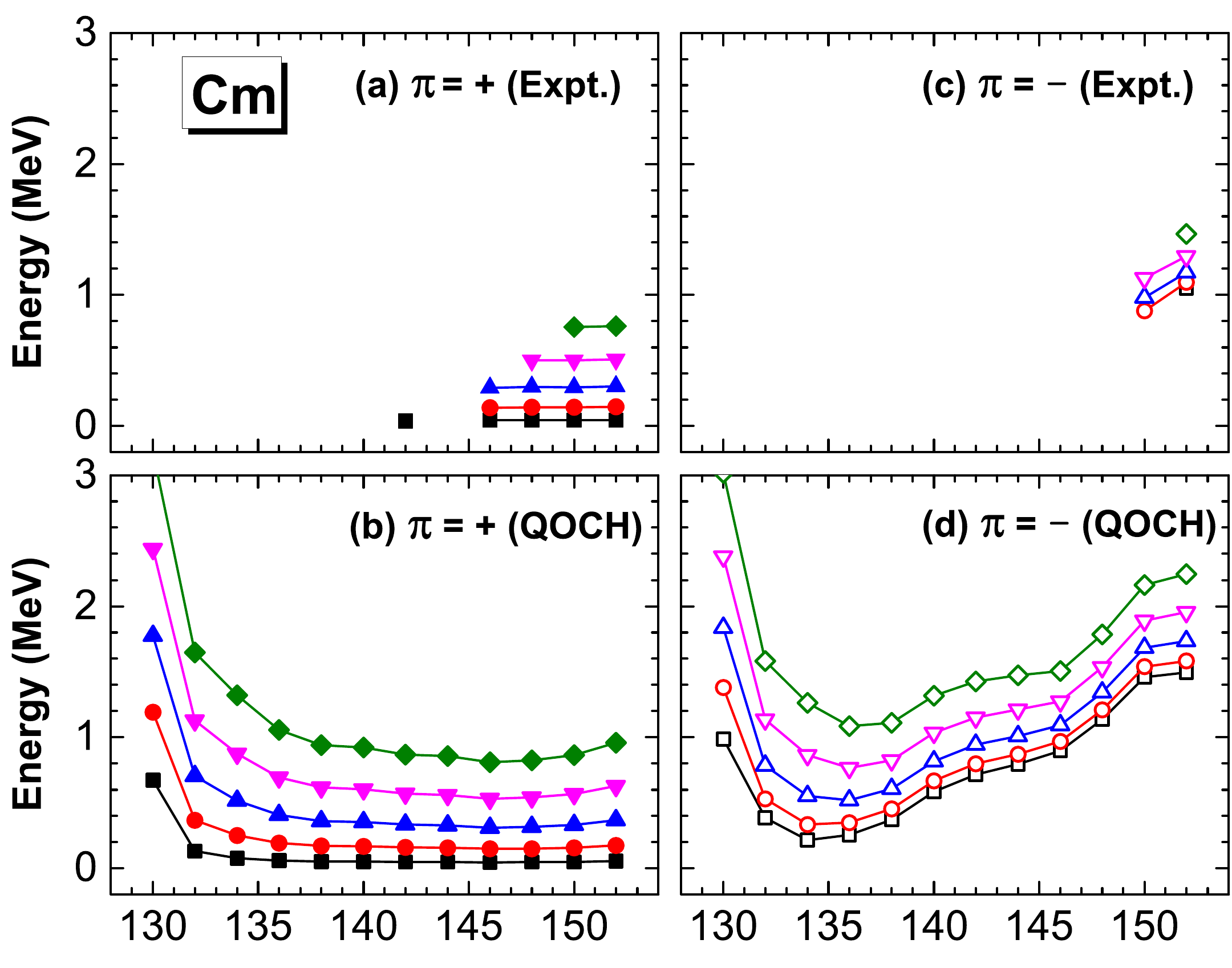}\vspace{-0.1cm}
\includegraphics[height=0.387\textwidth]{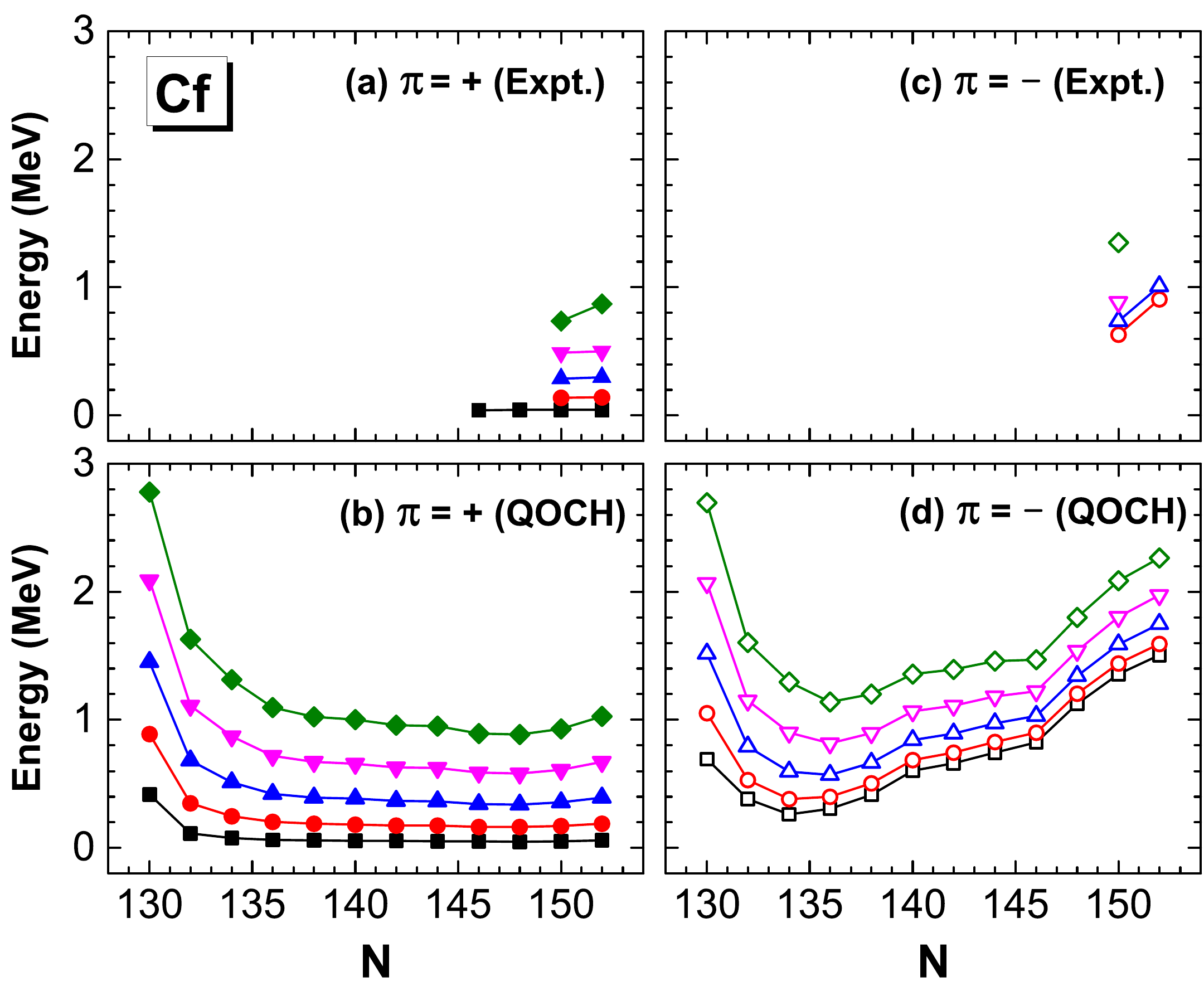}
\includegraphics[height=0.387\textwidth]{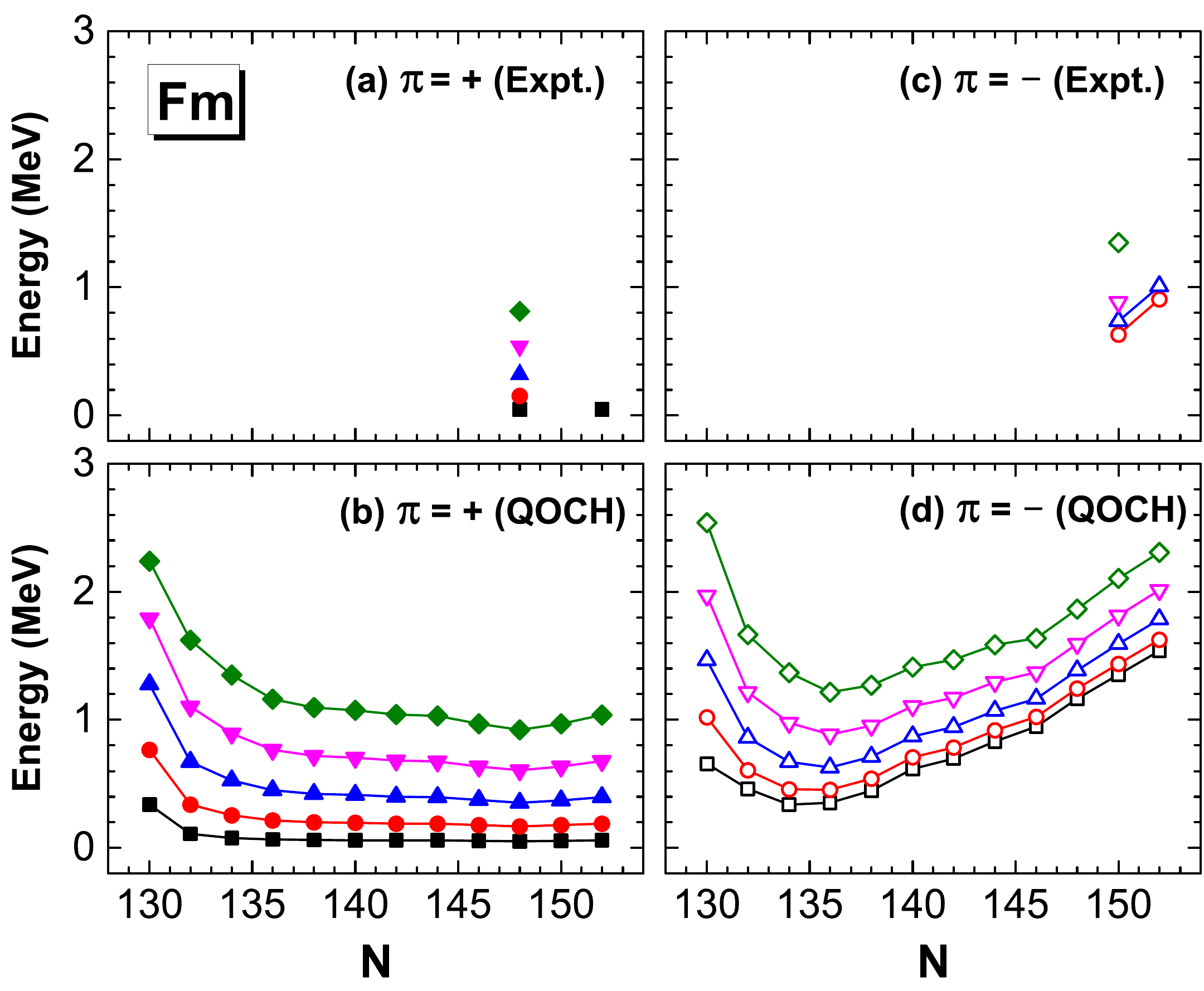}\vspace{-0.3cm}
\caption{(Color online) Same as in the caption to Fig. \ref{specLa} but for the Rn, Ra, Th, U, Pu, Cm, Cf, and Fm isotopes.}
\label{specAc}
\end{figure}

In Fig.~\ref{ExpTh} we illustrate the general quality of the QOCH model calculation based on the PC-PK1 functional and $\delta$-force pairing, by comparing the theoretical  excitation energies of low-lying states $2^+_1$, $4^+_1$, and $3^-_1$, and the $B(E2; 2^+_1\to0^+_1)$ values, to available data. The theoretical results are in reasonable agreement with experiment, both for the excitation energies and $E2$ transition rates, especially considering that the excitation spectra have been calculated in the lowest order approximation with the Inglis-Belyaev moments of inertia and perturbative cranking mass parameters.  Exceptions are found in some spherical nuclei, where the theoretical excitation energies are too high compared to the data (c.f. Figs. \ref{specLa} and \ref{specAc}). One of the reasons is that the present implementation of the QOCH model includes only 
quadrupole and octupole degrees of freedom with the additional constraint of axial symmetry. It does not take into account other, in some cases important, degrees of freedom such as two-quasiparticle configurations and triaxial deformations.

%\clearpage

%\clearpage

Figures \ref{specLa} and \ref{specAc} display the energy spectra of low-lying even-spin positive-parity states up to $J^\pi=10^+$, and the odd-spin negative-parity states up to $J^\pi=9^-$, for the fourteen isotopic chains.  The theoretical results are overall in reasonable agreement with the available data, except for the fact that the calculated spectra are too stretched in nuclei that are nearly spherical. For the positive-parity bands the excitation energies decrease rapidly up to $N\sim90$ and $N\sim136$ for the lighter and heavier mass regions, respectively, and then vary slowly with the addition of more neutrons. This is consistent with the evolution of the average quadrupole deformation $\langle\beta_2\rangle$ (cf. Fig. \ref{aveb2b3}). For the $K^\pi=0^-$ negative-parity bands in Fig. \ref{specLa}, the excitation energies in the lighter mass region decrease up to $N\sim88$, show little variation till $N\sim94$, and then increase gradually with neutron number. Similar results were also predicted by the interacting boson model (IBM) mapped from the mean-field potential energy surfaces using the Gongy interaction or relativistic EDFs \cite{Nomura14,Nomura15}. Deviations are found in the Sm and Gd isotopes with $N>90$, where the experimental excitation energies decrease along the isotopic chains whereas the theoretical values increase. This correlates with the diminishing of octupole fluctuations around the equilibrium minima of the DESs (cf. Fig. \ref{PESSmGd}). 

In the heavier mass region the excitation energies of the negative-parity bands exhibit a parabolic behaviour with minima at $N\sim136$. For the Ra and Th isotopes, minima are observed at $N\sim136$ for the experimental negative-parity states, whereas the functional PC-PK1 predict the minima at $N\sim138$. The theoretical minima gradually evolve to $N\sim134$ for the heavier isotopic chains. This is related to the onset of  octupole minima in the deformation energy surfaces (cf. Figs. \ref{PESRn} and \ref{PESPu}). The appearance of plateaus  in the calculated excitation energies of negative parity states at $N=142\sim146$ is attributed to the softness of the corresponding energy surfaces in the octupole direction, and is consistent with the available data for Th, U, and Pu isotopes.

\clearpage
\begin{figure}[ht]
\includegraphics[height=0.6\textwidth]{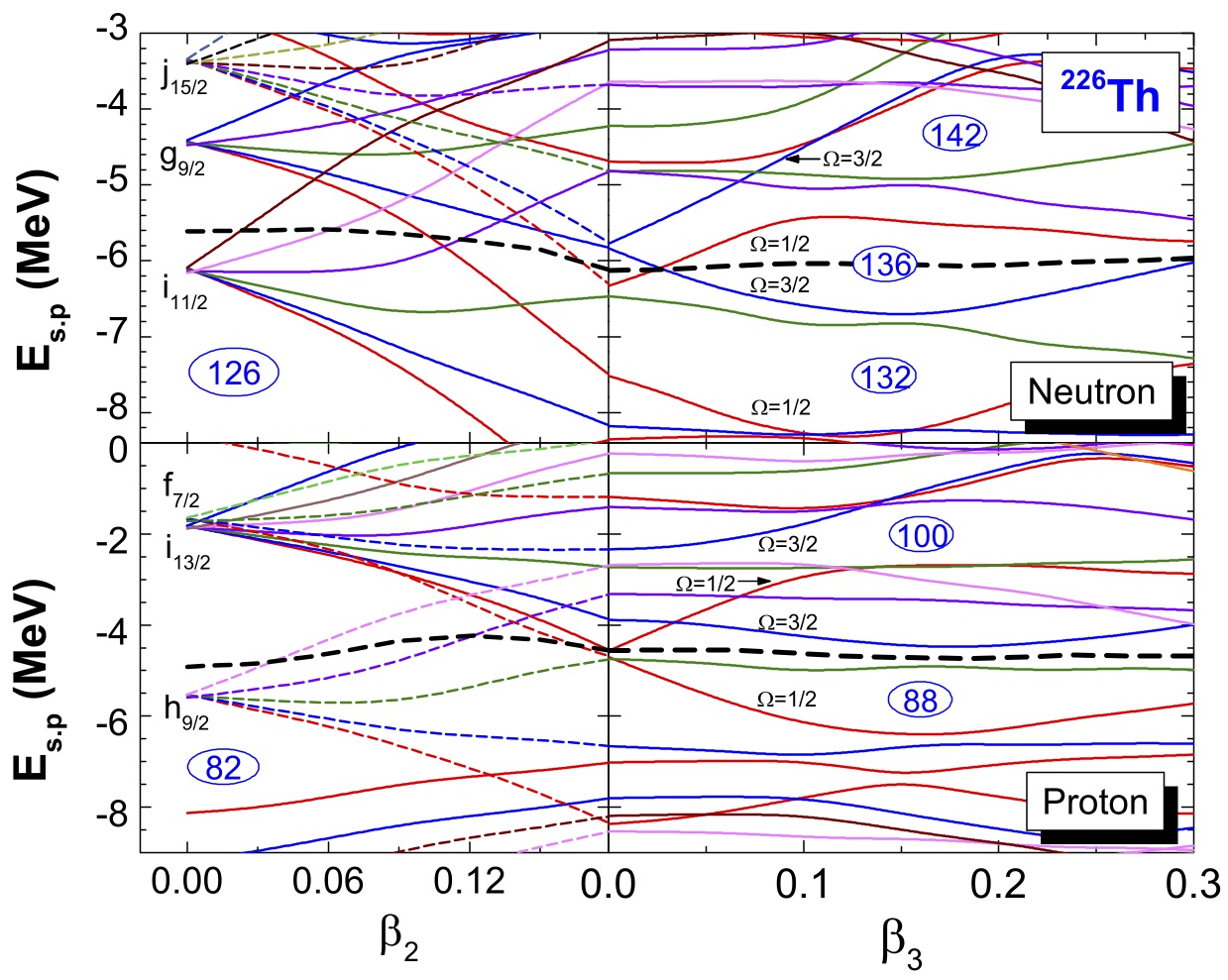}
\caption{(Color online) Single neutron and proton levels of $^{226}$Th as functions of deformation parameters. Each plot follows the quadrupole deformation parameter $\beta_2$ up to the position of the equilibrium minimum $\beta_2 = 0.18$, with the octupole deformation parameter kept constant at zero value. Then, for the constant value $\beta_2=0.18$, the path continues from $\beta_3 = 0$ to $\beta_3 = 0.3$. The thick dashed (black) curves denote the Fermi levels.}
\label{splevel}
\end{figure}

A microscopic picture of the onset of octupole deformation and octupole softness emerges from the dependence of the single-nucleon levels on the two deformation parameters. In Fig.~\ref{splevel} we plot the single neutron and proton levels of $^{226}$Th along a path in the $\beta_2-\beta_3$ plane. The path follows the quadrupole deformation parameter $\beta_2$ up to the position of the equilibrium minimum $\beta_2 = 0.18$, with the octupole deformation parameter kept constant at zero value. Then, for the constant value $\beta_2=0.18$, the path continues from $\beta_3 = 0$ to $\beta_3 = 0.3$. In the mean-field approach there is a close relation between the total binding energy and the level density around the Fermi level in the Nilsson diagram of single-particle energies. A lower-than-average density of single-particle levels around the Fermi energy results in extra binding, whereas a larger-than-average value reduces binding. Therefore, the onset of octupole minima around $^{226}$Th can be attributed to the low neutron-level density around the Fermi surface at $N\sim136$ and low proton-level density at $Z\sim90$. One notices the repulsion between the $\Omega=1/2$ pairs of levels that originate from the $(g_{9/2}, j_{15/2})$ spherical neutron levels and $(f_{7/2}, i_{13/2})$ spherical proton levels, respectively. 
A rather low neutron-level density is also predicted at $N\sim142$, induced by the repulsion between the $\Omega=3/2$ pair that originates from the 
$(g_{9/2}, j_{15/2})$ levels, which could cause the octupole softness in the isotopes with $N=142\sim146$. Moreover, a low proton-level density at $Z\sim100$, characterized by the repulsion between the $\Omega=3/2$ pair of levels originating from the $(f_{7/2}, i_{13/2})$ spherical levels, may lead to the onset of octupole deformation around $^{234}$Fm. We have also analyzed single-nucleon levels for the lighter mass region, and found that the onset of octupole minima around $^{144}$Ba can be attributed to the low neutron-level density around the Fermi surface at $N\sim88$ and low proton-level density at $Z\sim56$, characterized by the repulsion of the $\Omega=1/2$ pair from the $(f_{7/2}, i_{13/2})$ spherical neutron levels and the $(d_{5/2}, h_{11/2})$ spherical proton levels, respectively.

%\clearpage
\begin{figure}[ht]
\includegraphics[height=0.65\textwidth]{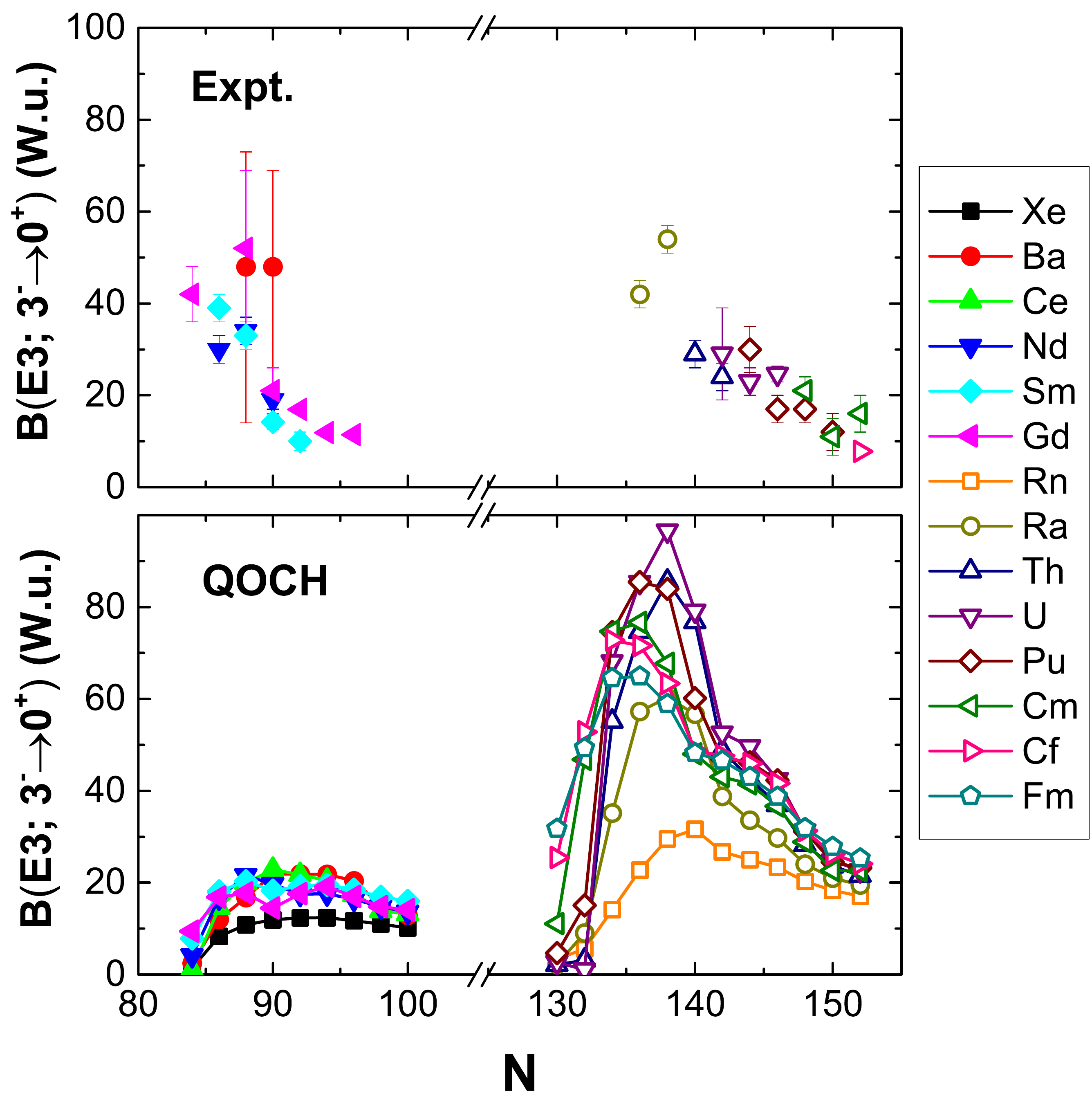}
\caption{(Color online) The experimental and theoretical $B(E3; 3^-_1\to 0^+_1)$ values as functions of neutron number for the fourteen isotopic chains analyzed in this study. The experimental values are from Refs. \cite{Kibedi02,Gaff13,Bucher16,Bucher17}.}
\label{BE3}
\end{figure}

In Fig.~\ref{BE3} we compare the experimental and theoretical $B(E3; 3^-_1\to 0^+_1)$ values as functions of neutron number for fourteen isotopic chains. The calculation generally reproduces the empirical values. Quantitatively, the QOCH underestimates the electric octupole transition rates of isotopes with $N\leqslant88$. This could be because the present model only includes the $K=0$ components (axial symmetry), and the mixing with $K\neq0$ might become important in nearly spherical nuclei \cite{Robledo11}. On the other hand, it appears that the QOCH overestimates the electric octupole transition rates in the heavier mass region. This indicates that the predicted octupole deformations and corresponding deformation energies are probably too large in this mass region, which is possibly related to weaker pairing correlations \cite{Agbe16}. Smaller values of the calculated $B(E3)$ for weakly deformed nuclei, and larger $B(E3)$ values for well-deformed nuclei have also been obtained in a systematic calculation based on the generator-coordinate extension of the Hartree-Fock-Bogoliubov self-consistent mean field theory with the  Gogny D1S effective interaction \cite{Robledo11}.

\begin{figure}[ht]
\includegraphics[height=0.9\textwidth]{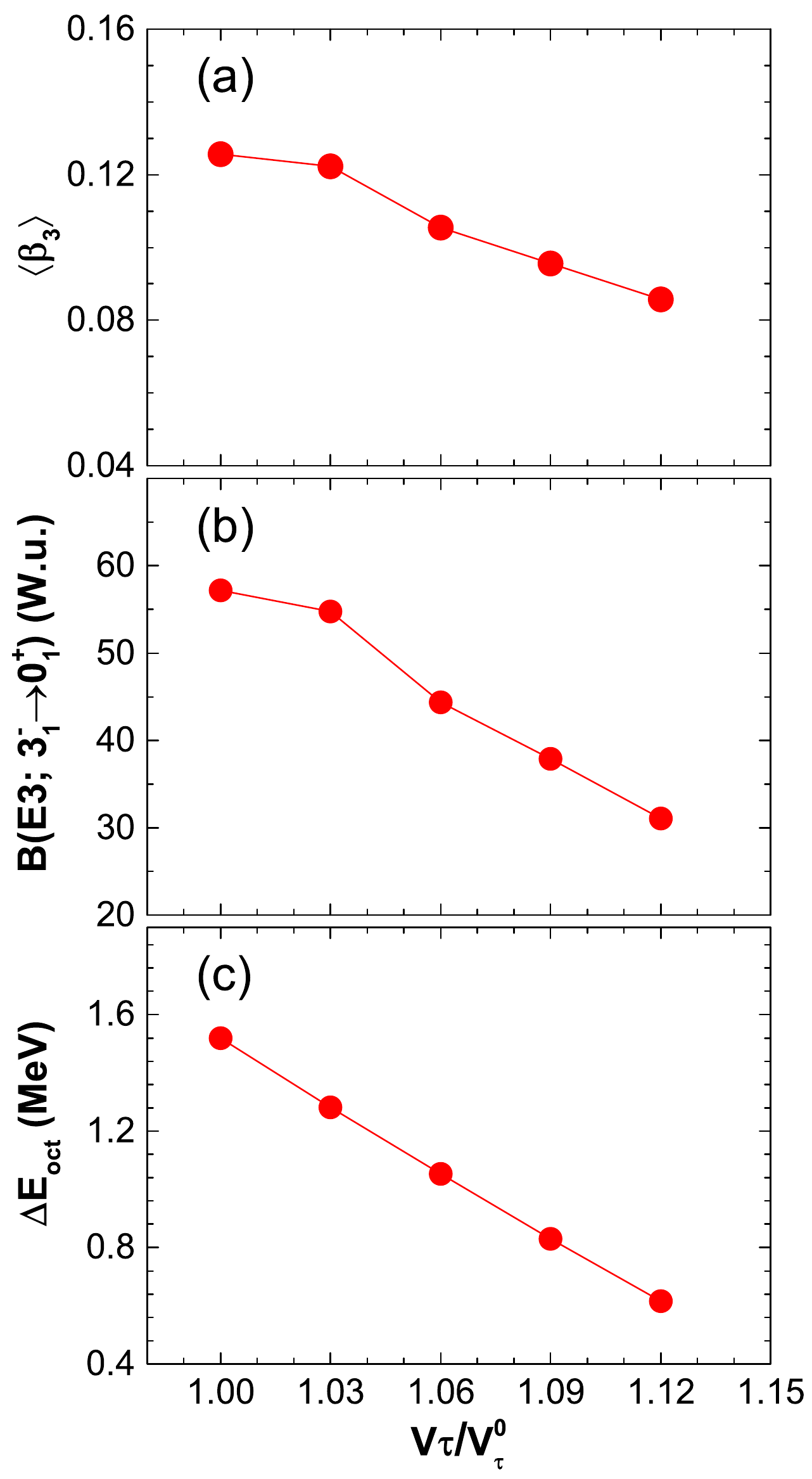}
\caption{(Color online) The calculated average octupole deformation $\langle\beta_3\rangle$ of the ground state (a), $B(E3; 3^-_1\to 0^+_1)$ (b), and octupole deformation energy $\Delta E_{\rm oct}$ (c), defined as the energy difference between the quadrupole deformed local minimum at $\beta_3=0$ and the global minimum, for $^{224}$Ra as functions of the ratio of the pairing strength $V_\tau$ to the original value $V^0_\tau$ used in this study ($\tau = n, p$).}
\label{be3-pair}
\end{figure}

To illustrate the relation between octupole deformation and pairing correlations, for $^{224}$Ra in Fig. \ref{be3-pair} we display the evolution of the calculated $\langle\beta_3\rangle$ of the ground state, the $B(E3; 3^-_1\to 0^+_1)$, and the octupole deformation energy $\Delta E_{\rm oct}$, as functions of the pairing strength $V_\tau$ ($\tau = n, p$). Note that all three quantities characteristic for octupole correlations exhibit a marked decrease as pairing strength increases.

\clearpage
\begin{figure}[ht]
\includegraphics[height=0.65\textwidth]{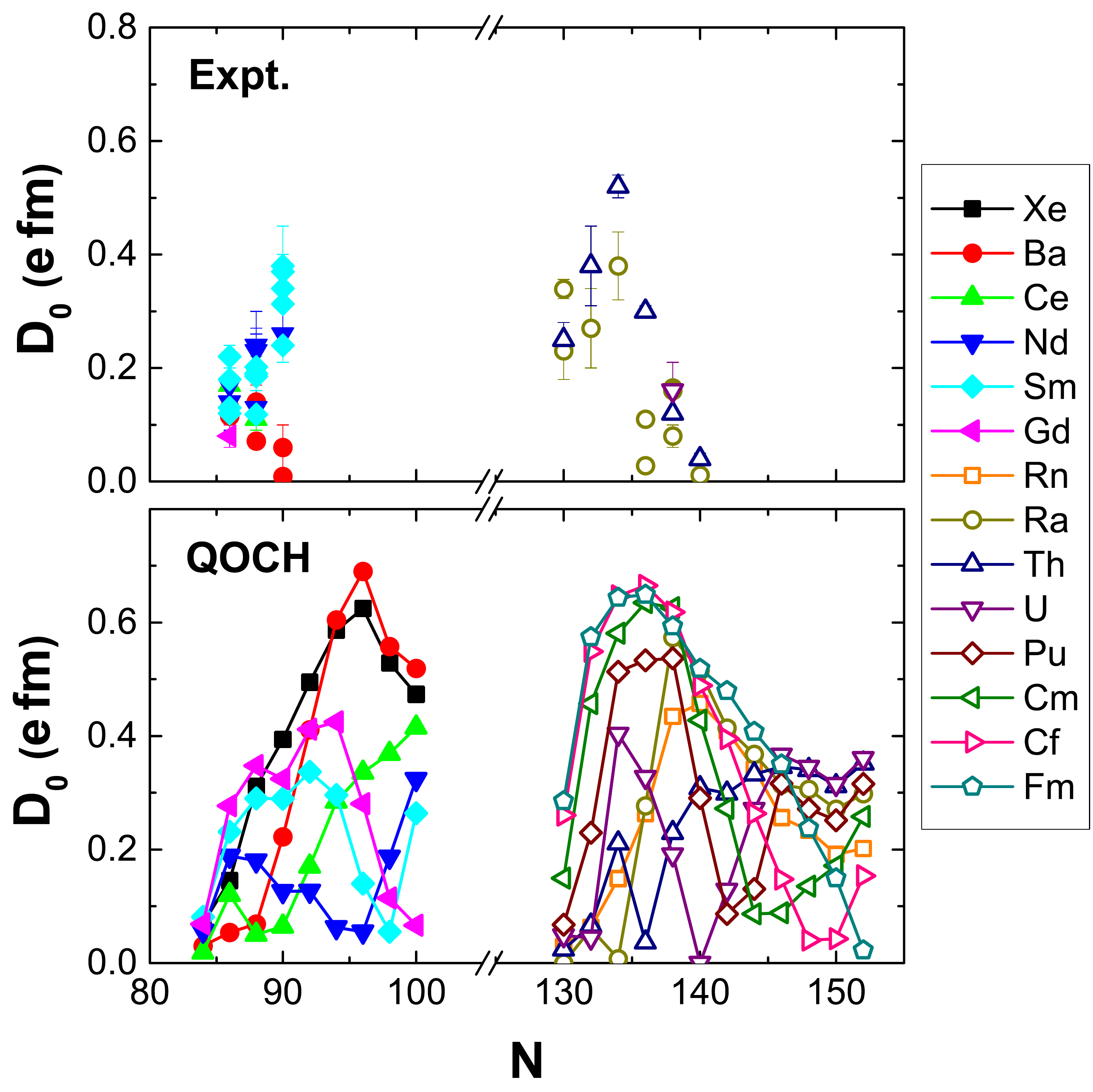}
\caption{(Color online) The experimental and theoretical electric dipole moments $D_0$ as functions of neutron number. The experimental values are from Ref. \cite{Butler96}.}
\label{D0}
\end{figure}

Finally, Fig.~\ref{D0} displays the experimental and theoretical intrinsic electric dipole moments $D_0$ as functions of neutron number for fourteen isotopic chains. The theoretical $D_0$ is calculated from the corresponding reduced transition probability using the relation 
\begin{equation}
B(E1; 1^-_1\to0^+_1)=\frac{3}{4\pi}D^2_0\langle 1010|00\rangle^2.
\end{equation}
We note that for one isotope there may be more than one experimental $D_0$, which correspond to different angular momenta. The theoretical results are 
of the same order of magnitude as the available data. However, large fluctuation with $Z$ and $N$ in the values of $D_0$ can occur due to shell effects and occupancy of different orbitals \cite{Butler96,Bucher17}, and the theoretical predictions fail to reproduce the detailed $N$ ($Z$) dependence of $D_0$.

%
%%%%%%%%%%%%%%%%%%%%%%%%%%%%%%%%%%%%%%%%%%%%%%%%%%
\section{\label{IV}Summary}

In the present study we have performed a microscopic analysis of octupole shape transitions in fourteen isotopic chains characteristic for two regions of octupole deformation and collectivity: Xe, Ba, Ce, Nd, Sm, Gd, Rn, Ra, Th, U, Pu, Cm, Cf, and Fm.  Starting from self-consistent binding energy maps in the $\beta_2-\beta_3$ plane, calculated with the RMF+BCS model based on the functional PC-PK1 and $\delta$-interaction pairing, a recent implementation of the quadrupole-octupole collective Hamiltonian for  vibrations and rotations has been used to calculate the excitation spectra and transition rates of 150 even-even nuclei. The parameters that determine the collective Hamiltonian: the vibrational inertial functions, the moments of inertia, and the zero-point energy corrections, are calculated using single-quasiparticle energies and wave functions that correspond to each point on the self-consistent RMB+BCS deformation energy surface of a given nucleus. The diagonalization of the collective Hamiltonian yields the excitation energies and wave functions used to calculate various observables.

The microscopic deformation energy surfaces exhibit characteristic transitions with increasing neutron number: from spherical quadrupole vibrational to stable octupole deformed nuclei, to octupole vibrations typical for $\beta_3$-soft potentials, and finally to well-deformed prolate shapes in the Ba, Ce, Ra, Th, U, Pu, Cm, Cf, and Fm isotopic chains. For Nd, Sm, and Gd isotopes one finds soft energy surfaces with respect to both quadrupole and octupole degrees of freedom in transitional nuclei with $N\sim90$. 
The systematics of the energy spectra and transition rates, associated with both positive- and negative-parity yrast states, points to the appearance of prominent octupole correlations around $N\sim90$ and $N\sim136$, and the corresponding lowering in energy of negative-parity bands with respect to the positive-parity ground-state band. The energy plateaus of negative-parity bands at $N=142\sim146$ can be related to the soft energy surfaces in octupole direction in the heavier mass region. In addition, the spectroscopic properties predicted by the QOCH model are generally in reasonable agreement with the systematics of available data, and consistent with previous GCM calculations and IBM results. Differences between empirical properties and theoretical results, especially for the electric octupole transition rates  and electric dipole moments, indicate that the theoretical  framework necessitates further development, such as the inclusion of the triaxial degree of freedom and two-quasiparticle configurations. 
%
%
%----------------------------------------------------------------------------------------------------------------
\begin{acknowledgements}
We thank Z. Xu for helpful discussions. This work was supported in part by the NSFC under Grant Nos. 11475140 and 11575148,
the Croatian Science Foundation -- project ``Structure and Dynamics
of Exotic Femtosystems" (IP-2014-09-9159), the QuantiXLie Centre of Excellence,
and the Chinese-Croatian project ``Microscopic Energy Density Functionals Theory for Nuclear Fission''.
\end{acknowledgements}
%\clearpage
%====================================================%
%

\end{document}